\documentclass[twocolumn%
,secnumarabic%
,amssymb,aps,pra,nobibnotes,longbibliography,superscriptaddress]{revtex4-1}
\usepackage{amsmath}
\usepackage{graphicx}
\usepackage{float}
\usepackage{color}
\AtBeginDocument{%
    \newwrite\bibnotes
    \def\bibnotesext{Notes.bib}
    \immediate\openout\bibnotes=\jobname\bibnotesext
    \immediate\write\bibnotes{@CONTROL{REVTEX41Control}}
    \immediate\write\bibnotes{@CONTROL{%
    apsrev41Control,author="08",editor="1",pages="1",title="0",year="1"}}
     \if@filesw
     \immediate\write\@auxout{\string\citation{apsrev41Control}}%
    \fi
}%

\expandafter\ifx\csname package@font\endcsname\relax\else
 \expandafter\expandafter
 \expandafter\usepackage
 \expandafter\expandafter
 \expandafter{\csname package@font\endcsname}%
\fi
\DeclareRobustCommand\substyle{\name@idx{document substyle}}%
\DeclareRobustCommand\classoption{\name@idx{document class option}}%
\DeclareRobustCommand\classname{\name@idx{document class}}%
\def\name@idx#1#2{%
 {\ttfamily#2}%
 \index{#2\space#1=\string\ttt{#2}\space#1}\index{#1>#2=\string\ttt{#2}}%
}%

\DeclareUnicodeCharacter{2212}{-}

\begin{document}
\title{Unified Framework for Charge-Spin Interconversion in Spin-Orbit Materials}
\author{Shehrin Sayed}\email{ssayed@berkeley.edu}
\affiliation{Electrical Engineering and Computer Sciences, University of California, Berkeley, CA 94720, USA}
\affiliation{Materials Sciences Division, Lawrence Berkeley National Laboratory, Berkeley, CA 94720, USA}
\author{Seokmin Hong}
\affiliation{Center for Spintronics, Korea Institute of Science and Technology, Seoul 02792, Republic of Korea.}\email{shong@kist.re.kr}
\author{Xiaoxi Huang}
\affiliation{Materials Science and Engineering,
	University of California, Berkeley, CA 94720, USA}
\author{Lucas Caretta}
\affiliation{Materials Science and Engineering,
	University of California, Berkeley, CA 94720, USA}
\affiliation{Materials Sciences Division, Lawrence Berkeley National Laboratory, Berkeley, CA 94720, USA}
\author{Arnoud S. Everhardt}
\affiliation{Materials Science and Engineering,
	University of California, Berkeley, CA 94720, USA}
\author{{Ramamoorthy Ramesh}}
\affiliation{Materials Science and Engineering,
	University of California, Berkeley, CA 94720, USA}
\affiliation{Materials Sciences Division, Lawrence Berkeley National Laboratory, Berkeley, CA 94720, USA}
\author{{Sayeef Salahuddin}}
\affiliation{Electrical Engineering and Computer Sciences,
	University of California, Berkeley, CA 94720, USA}
\affiliation{Materials Sciences Division, Lawrence Berkeley National Laboratory, Berkeley, CA 94720, USA}
\author{{Supriyo Datta}}
\email{datta@purdue.edu}
\affiliation{Electrical and Computer Engineering, Purdue University, West Lafayette, IN 47907, USA}

\begin{abstract}
Materials with spin-orbit coupling are of great interest for various spintronics applications due to the efficient electrical generation and detection of spin-polarized electrons. Over the past decade, many materials have been studied, including topological insulators, transition metals, Kondo insulators, semimetals, semiconductors, and oxides; however, there is no unifying physical framework for understanding the physics and therefore designing a material system and devices with the desired properties. We present a model that binds together the experimental data observed on the wide variety of materials in a unified manner. We show that in a material with a given spin-momentum locking, the density of states plays a crucial role in determining the charge-spin interconversion efficiency, and a simple inverse relationship can be obtained. Remarkably, experimental data obtained over the last decade on many different materials closely follow such an inverse relationship. We further deduce two figure-of-merits of great current interest: the spin-orbit torque (SOT) efficiency (for the direct effect) and the inverse Rashba-Edelstein effect length (for the inverse effect), which statistically show good agreement with the existing experimental data on wide varieties of materials. Especially, we identify a scaling law for the SOT efficiency with respect to the carrier concentration in the sample, which agrees with existing data. Such an agreement is intriguing since our transport model includes only Fermi surface contributions and fundamentally different from the conventional views of the SOT efficiency that includes contributions from all the occupied states.
\end{abstract}

\maketitle

\section{Introduction}

Recently, charge-spin interconversion in various materials exhibiting spin-momentum locking (SML), e.g., topological insulators (TI) \cite{JonkerNatNano2014,KLWangNanoLett2014, DasNanoLett2015, SamarthPRB2015, YPChenSciRep2015, Samarth_PRB_2015, YoichiPRB2016}, semiconductors \cite{Koo_New,Koo_JAppPhys_2012}, transition metals \cite{PhysRevB.88.064414,Parkin_NPhys_2015,doi:10.1063/1.4922084,Ralph_Science_2012,Ralph_APL_2012,JimmySciRep2019,PhysRevLett.106.126601,PhysRevLett.109.156602,PhysRevB.89.054401}, semimetals \cite{Li_NatComm_2018}, oxides \cite{FertLAOSTO2016,PhysRevApplied.12.034004,Arnoud_PRMat_2019,Nan16186}, and antiferromagnets \cite{Ohno_NatMat_2016}, superconductor \cite{Gotlieb1271}, are growing interest for efficient spintronic applications. The sheer variety of materials is exciting, but it also poses a daunting challenge in understanding the underlying physics and designing devices for various applications with desired properties.

In this paper, we present a model (see Fig. \ref{fig_circuit}) that binds together the charge-spin interconversion observed on the wide variety of materials in a unified manner using only four model parameters: number of modes or density of states, the strength of SML, mean free path, and interface spin conductance. These four parameters are independently measurable and theoretically well-understood. We show that, in a material with a given SML, the number of modes in the channel, which is related to the material density of states around the Fermi energy, plays a crucial role in determining the charge-spin interconversion efficiency. We point out a simple inverse relationship between the interconversion efficiency and the material density of states. Remarkably, experimental data obtained over the last decade on many different material systems closely follow this inverse relationship (see Fig. \ref{spin_sig}). We provide an intuitive understanding of the origin of such a scaling trend.

We discuss several figure-of-merits of great interest in emerging applications. We show that the charge current induced spin voltage and its reciprocal effect, which are of great interest to read out magnetization states in logic \cite{MESO2015} and memory \cite{SayedEDL2017} applications, can be enhanced by lowering the density of states in a given material. We quantify the SML strength in various materials based on the available experimental data and compare it with the theoretical expectations from various origins of the SML. We further relate the four model parameters to widely used figure-of-merits: the spin-orbit torque (SOT) efficiency (see Figs. \ref{fig_sha} and \ref{fig_sio}) and the inverse Rashba-Edelstein effect (IREE) (see Fig. \ref{fig_IREE}), which show good agreement with existing experiments. We show that the figure-of-merit for the spin voltage multiplied by the measured interface spin conductance yields the SOT efficiency and also scales inversely with the number of modes in the channel. We identify a scaling law with the carrier concentration ($n$) in the sample that the SOT efficiency $\propto n^{-\frac{2}{3}}$. Such a trend can be useful for future device design because the carrier concentration in various materials can be controlled with doping, electric gating, and strain modulation.

We derive the model from a semiclassical equation \cite{Sayed_PRAppl_2018} that we previously obtained from the Boltzmann transport equation using four electrochemical potentials based on the sign of the group velocity and the spin index. Within our model, the transport parameters are described with electronic states around the Fermi energy, which is different from conventional views. Conventional models using the spin Berry curvature include contributions from all
occupied states. By contrast, our model includes only Fermi surface contributions
making it fundamentally different and this could help identify interesting results as the field evolves. For example, we discuss an interesting observation in recent experiments on correlated oxides \cite{Arnoud_PRMat_2019, Nan16186} where the SOT efficiency decreased with the sample resistivity, contrary to the conventional view that the SOT efficiency increases in a resistive sample. We show that the experimental results closely follow the scaling trend of our model with respect to the measured carrier concentration. We briefly discuss the parameter conditions required to observe SOT efficiencies $>1$ within our model, which is of great current interest. In the present paper, we restrict our discussion to the cases where the SOT efficiencies are $<1$.

The paper is organized as follows. In Section II, we present a resistance matrix model and describe its four key parameters. In Section III, we discuss charge current induced spin voltage and show that the spin voltage scales inversely with the channel number of modes or material density of states over a broad range of materials. We extract the degree of spin-momentum locking in diverse materials and extend the spin voltage model to derive an SOT efficiency model, which is in good agreement with the experiments on various materials. We also compare our model with conventional views. In Section IV, we discuss spin current induced charge voltage and extend the model to derive the inverse Rashba-Edelstein effect length, which shows good agreement with existing experiments on diverse materials. We end with a summary in Section V.

\section{Resistance Matrix Model}

\begin{figure}
	%\vspace*{1cm}
	\includegraphics[width=0.45\textwidth]{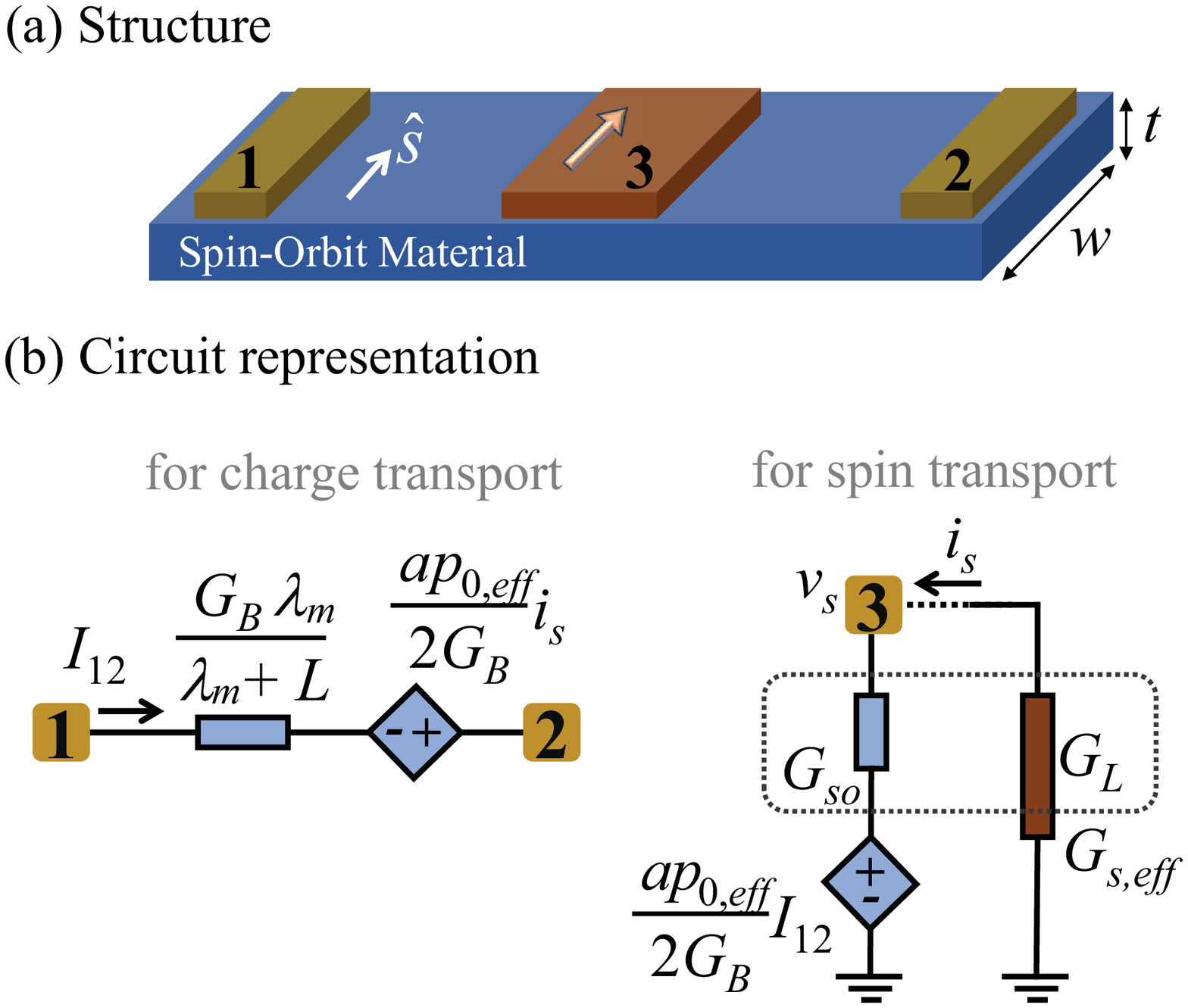}
	\centering
	\caption{(a) A three-terminal based structure with a spin-orbit material. (b) Circuit representation for charge to spin conversion in a spin-orbit material.}\label{fig_circuit}
\end{figure}

\begin{figure*}
	%\vspace*{1cm}
	\includegraphics[width=1.0 \textwidth]{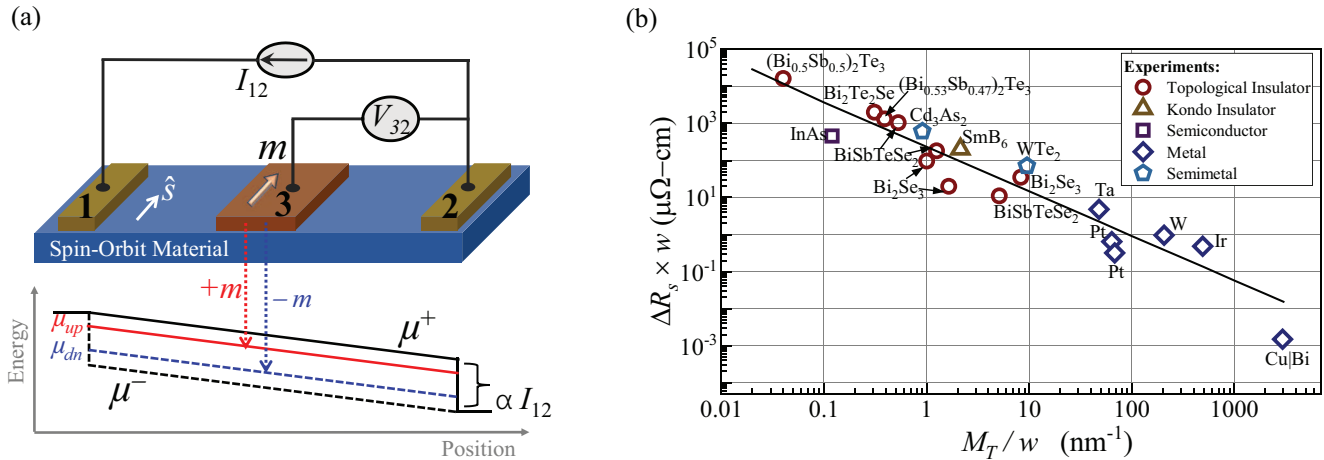}
	\centering
	\caption{(a) Setup to measure
charge current induced spin voltage in a spin-orbit material using a ferromagnetic contact with magnetization $m$. (b) Charge-spin interconversion resistance, $\Delta {R_S}$, in diverse classes of materials as a function of channel number of modes ($M_T$), which is related to the material density of states near the Fermi energy. Experimental data points include: topological insulators ((Bi$_{0.5}$Sb$_{0.5}$)$_2$Te$_3$ \cite{SamarthPRB2015}, (Bi$_{0.53}$Sb$_{0.47}$)$_2$Te$_3$ \cite{KLWangNanoLett2014}, Bi$_{2}$Se$_{3}$ \cite{vanWees_PRB_2015, DasNanoLett2015, SamarthPRB2015}, Bi$_2$Te$_2$Se \cite{YPChenSciRep2015}, and BiSbTeSe$_2$ \cite{Ando_arXiv_2016}), transition metals (Ta \cite{Liu_NPhys_2014}, Pt \cite{Pham_NanoLett_2016,Liu_NPhys_2014}, W \cite{Liu_NPhys_2014}, and Ir \cite{Liu_NPhys_2014}), metallic interface (Cu$|$Bi \cite{IssaCuBi2016}), narrow bandgap semiconductor (InAs \cite{Koo_JAppPhys_2012, Koo_New}), topological Kondo insulator (SmB$_6$ \cite{Hong_KIST_arXiv_2018}), and semimetals (WTe$_2$ \cite{Li_NatComm_2018} and Cd$_3$As$_2$ \cite{PhysRevLett.124.116802}). Solid black line represents the theoretical trend.}\label{spin_sig}
\end{figure*}

Our discussions in this paper is based on the following model for an arbitrary channel with SML
\begin{equation}
\label{R_Mat}
\left\{ {\begin{array}{*{20}{c}}
{{V_1} - {V_2}}\\
{{v_s}}
\end{array}} \right\} = {\rm{ }}\left[ {\begin{array}{*{20}{c}}
{\dfrac{{\lambda_m  + L}}{{{G_B}\lambda_m }}}&{ - \dfrac{{\alpha {p_{0,eff}}}}{{2{G_B}}}}\\\\
{\dfrac{{\alpha {p_{0,eff}}}}{{2{G_B}}}}&{\dfrac{1}{{{G_{so}}}}}
\end{array}} \right]\left\{ {\begin{array}{*{20}{c}}
{{I_{12}}}\\
{{i_s}}
\end{array}} \right\},
\end{equation}
obtained from a semiclassical equation \cite{Sayed_PRAppl_2018} in steady-state, for an uniform structure with uniform spin potential and elastic scatterings in the channel, see Appendix \ref{AppA}. Here $V_{1,2}$ are the charge voltages on terminals 1 and 2, $I_{12}$ is the charge current flowing in the channel, $v_s$ is the spin voltage in the SML channel, $i_s$ is the spin current injected from the SML channel to the adjacent layer, $L$ is the channel length, and $\alpha=2/\pi$ is an angular averaging factor for the spin distribution in the channel. We represent Eq. \eqref{R_Mat} into a circuit in Fig. \ref{fig_circuit}.  Such a representation will be useful for a straightforward analysis of emerging devices and circuits implemented using spin-orbit materials.

A resistance matrix similar to Eq. \eqref{R_Mat} was reported in Ref. \cite{Sayed_PRAppl_2018} for structures with non-magnetic potentiometric (non-invasive) contacts on SML channels, which was derived under specific assumptions on the scattering mechanisms in the channel. Eq. \eqref{R_Mat} in this manuscript applies to a structure with general scattering processes and arbitrary magnetic or non-magnetic contacts. Although the resistance matrix is Eq. \eqref{R_Mat} is similar to Ref. \cite{Sayed_PRAppl_2018}, the effective model parameters are different and considers various possible origins of the SML.

The first diagonal term in Eq. \eqref{R_Mat} represents the charge resistance of the channel and the second diagonal term represents the spin source resistance that limits the spin current injection into the adjacent layer. The off-diagonal terms indicate the coefficients for charge-spin interconversion. The model in Eq. \eqref{R_Mat} describes spin-dependent transport phenomena in terms of only four model parameters representing the channel: 
\begin{enumerate}
    \item $G_B=(q^2/h)M_T$ is a conductance parameter proportional to the total number of modes, $M_T$, in the channel ($q$ is the electron charge and $h$ is the Planck's constant),
    \item $p_{0,eff}$ is the effective strength of the SML,
    \item $\lambda_m$ is the effective mean free path, and
    \item $G_{so}$ is the source conductance of the spin transport in the channel that limits the spin current extracted from the channel by a highly spin conductive load.
\end{enumerate}
The details of the electronic structure of the materials, growth conditions, and different physical structures and interfaces could result in different effective values of the parameters; however, the general scaling law presented in the manuscript holds and statistically show a good agreement with a large number of experimental data. The parameters are described in detail below.

\subsection{Channel number of modes and the material density of states}
The conductance parameter $G_B$ in Eq. \eqref{R_Mat} represents the total number of modes, $M_T$, in the channel, given by
\begin{equation}
\label{MT_Dv}
    G_B = \dfrac{q^2}{h} M_T.
\end{equation}
The total number of modes in the channel is related to the density of states around the Fermi energy per unit length of the channel, $D$, and the group velocity of the electronic states, $v$, as given by \cite{Datta_LNE_2012}
\begin{equation}
\label{MT_Dv}
    M_T = \frac{hDv}{2}\eta,
\end{equation}
where $\eta$ is a geometrical factor, which is $1$, $\frac{2}{\pi}$, and $\frac{1}{2}$ in a 1D, 2D, and 3D channel, respectively. $M_T$ represents the number of electron wavelengths that fit into the channel cross-section \cite{Datta_LNE_2012} and related to the carrier concentration in the sample as
\begin{subequations}
\label{Mt_n}
	\begin{equation}
	\label{Mt_n_3d}
		\frac{M_T}{wt}=\sqrt[3]{\dfrac{9\pi}{8}}\,n^\frac{2}{3},\;\;\;\;\;\text{(for a 3D channel)}
	\end{equation}
	\begin{equation}
		\frac{M_T}{w}=\sqrt{\dfrac{2}{\pi}}\,n_s^\frac{1}{2},\;\;\;\;\;\;\text{(for a 2D channel)}
	\end{equation}
\end{subequations}
where $n_s$ and $n$ are the carrier concentrations in a 2D and a 3D channel, respectively, $w$ is the width, and $t$ is the thickness of the channel. See Appendix \ref{AppC}. 

%Note that Eq. \eqref{R_Mat} describes charge-spin interconversion in a conductor with a finite density of states around the Fermi energy, i.e., a finite $M_T$. $M_T=0$ indicates that either $D=0$ (i.e., an insulator)  or $v=0$ (i.e., electrons are not mobile), thus not a physical number for Eq. \eqref{R_Mat}.

\subsection{Strength of the spin-momentum locking}
\label{sec_SML}

The parameter $p_{0,eff}$ in our model quantifies the strength of SML in an arbitrary channel. $p_{0,eff}$ has two components:
\begin{equation*}
    p_{0,eff} = \epsilon p_0 + p_{rs},
\end{equation*}
where $\epsilon$ a unit less parameter that depends on the contact conductance, contact polarization, and scattering rates in the channel. The component proportional to $p_{rs}$ is induced by impurities that can cause a difference in the scattering rates between up and down polarized states, which is explained in details in Appendix \ref{AppA}.

The component proportional to $p_0$ is induced from the band structure of the material, which is given by \cite{Sayed_PRAppl_2018}
\begin{equation}
\label{deg_SML}
    p_0=\dfrac{M-N}{M+N},
\end{equation}
where $M$ is the number of modes for the up (and down) spin-polarized electronic states with the positive (and the negative) group velocity, see Appendix \ref{AppA}. Similarly, $N$ is the number of modes for the down (and up) spin-polarized electronic states with the positive (and the negative) group velocity. The total number of modes $M_T=M+N$. The parameter $p_0$ varies between 0 and 1. $p_0=0$ represents a normal metal, where up and down spin-polarized states have the same number of modes in the channel for positive and negative group velocities (i.e., $M=N$) and the channel exhibits no SML. $p_0=1$ (i.e., $N=0$) represents perfect SML where all electronic states with positive group velocity are up spin-polarized, and all electronic states with negative group velocity are down spin-polarized. Such a perfect SML is observed for an ideal topological band insulator.

For a general SML channel, the value of $p_0$ can be any value between 0 and 1 and depends on the underlying band structure and origin of the SML. For example, for a Rashba channel, Eq. \eqref{deg_SML} becomes (see Appendix \ref{AppB})
\begin{equation}
\label{RashbaSML}
p_0 = \dfrac{{{\alpha_R}}}{{\sqrt {\alpha_R^2 + \left(\hbar v_F\right)^2}}},
\end{equation}
where $\alpha_R$ is the Rashba coefficient, $v_F$ is the Fermi velocity, and $\hbar=h/(2\pi)$ is the reduced Planck constant. For weak Rashba channels, $p_0\approx{\alpha_R}/(\hbar{v_F})\;\ll\;1$.

Note that $p_{0,eff}$ can be non-zero even in a normal metal ($p_0=0$) due to a contribution from $p_{rs}$, if we can introduce impurities that can induce a difference in the scattering rates between up and down spin-polarized states, as described in Appendix \ref{AppA}. In this manuscript, we do not assume any particular origin of $p_{0,eff}$. We keep the parameter $p_{0,eff}$ general and extract it from existing experiments and comment on its strength based on results on known systems. We show that for a given $p_{0,eff}$ how the density of states or number of modes plays a crucial role in determining the efficiency of charge-spin interconversion.

\subsection{Mean free path}
The mean free path, $\lambda_m$, refers to the backscattering length of electrons, which determines the charge resistance of the sample. The mean free path and the total number of modes in the channel ($M_T$) are related to the measured resistivity ($\rho$) / sheet resistance ($R_{sheet}$) as
\begin{equation}
\label{Mt_rho}
    \frac{\rho }{{wt}}=\frac{{{R_{sheet}}}}{w} = \frac{h}{{{q^2 M_T}\lambda_m }}.
\end{equation}

\subsection{Spin source conductance}
Spin-orbit materials generate non-equilibrium spins, which can be injected into an adjacent layer. The internal spin source conductance, $G_{so}$, limits the spin current injected into a highly spin conductive layer, similar to the internal source resistance of a battery. In our model, $G_{so}$ is given by
\begin{equation}
\label{source_cond}
    G_{so} = \dfrac{4G_B\,L_f}{\alpha^2\left(1-p_0^2\right)\lambda_{s0}},
\end{equation}
see Appendix \ref{AppA} for details. Here, $\lambda_{s0}$ is a spin-dependent scattering length, and $L_f$ is the length of the ferromagnet.  Eq. \eqref{source_cond} (a component of Eq. \eqref{R_Mat}) was derived under the assumption of elastic scatterings, uniform structure, and uniform spin potential in the channel. The assumptions can be revisited as the field evolves, which may result in an effective change in the values of the model parameters; however, the general structure of Eq. \eqref{R_Mat} will not change.

In the presence of an adjacent layer with spin conductance $G_L$, see Fig. \ref{fig_circuit}, the effective conductance, $G_{s,eff}$, that determines the injected spin current is given by 
\begin{equation}
\label{eff_source_cond}
    G_{s,eff} = \dfrac{G_{so} G_L}{G_{so}+G_L}.
\end{equation}
For an adjacent ferromagnetic contact, $G_{L}$ is modeled with the bare interface spin mixing conductance \cite{Camsari_SciRep_2015, BauerPRB2013}. When the magnetization $\vec m$ is perpendicular to the spin polarization $\hat s$, $G_L$ is the real part of the spin mixing conductance $G_{r}^{\uparrow\downarrow}$  \cite{Camsari_SciRep_2015, BauerPRB2013}, and $G_{s,eff}$ is typically expressed in terms of an effective spin mixing conductance, $g_{r,eff}^{\uparrow\downarrow}$, in the units of m$^{-2}$, as given by
\begin{equation}
\label{eff_source_cond_perp}
    G_{s,eff} = \dfrac{G_{so} G_{r}^{\uparrow\downarrow}}{G_{so}+G_{r}^{\uparrow\downarrow}} = \frac{2q^2}{h} g_{r,eff}^{\uparrow\downarrow}w_f L_f,
\end{equation}
where $g_{r,eff}^{\uparrow\downarrow}$ can be measured from ferromagnetic resonance experiments on the bilayer \cite{Mellnik2014a,PhysRevLett.104.046601,Yang_APL_2013}. Here, $w_f$ and $L_f$ are width and length of the FM layer.

\begin{table*}
%		\captionsetup{justification=centering}
\scriptsize
\caption{Charge-Spin Interconversion Resistance in Diverse Materials.}\label{tab_Exp_Stength}
\begin{center}
	\begin{tabular}{||c | c | c | c |c| c||} 
		\hline
		\label{delRs_table}
		SOC Material & Ferromagnet ($p_f$) & $\Delta R_S$ ($\Omega$) & $w$ ($\mu$m) & $t$ (nm) & $\Delta R_S \times w$ ($\mu\Omega$-cm)\\ [0.5ex] 
		\hline\hline
		(Bi$_{0.5}$Sb$_{0.5}$)$_2$Te$_3$ \cite{SamarthPRB2015} & CoFeB $|$ MgO (0.5) & 20 & 8 & 7 & 16000 \\ 
		\hline
		(Bi$_{0.53}$Sb$_{0.47}$)$_2$Te$_3$ \cite{KLWangNanoLett2014} & Co $|$ Al$_2$O$_3$ (0.42) & 5.2 & 2.5 & 9 & 1300 \\ 
		\hline
		Bi$_{2}$Se$_{3}$ \cite{vanWees_PRB_2015} & Co $|$ TiO$_2$ (0.3) & 0.2 & 1 & 20 & 20 \\ 
		\hline
		Bi$_{2}$Se$_{3}$ \cite{DasNanoLett2015} & Co $|$ TiO$_2$ (0.2) & 0.07 & 5 & 40 & 35 (dev-1)\\ 
		\hline
		Bi$_{2}$Se$_{3}$ \cite{SamarthPRB2015} & CoFeB $|$ MgO (0.5) & 0.12 & 8 & 7 & 96 \\ 
		\hline
		Bi$_2$Te$_2$Se \cite{YPChenSciRep2015} & NiFe $|$ Al$_2$O$_3$ (0.45) & 2.2 & 9 & 10 & 1980 \\ 
		\hline
		BiSbTeSe$_2$ \cite{Ando_arXiv_2016} & NiFe $|$ MgO (0.45) & 0.013 & 8.5 & 172 & 11 (dev-1)\\ &&0.4&4.5&82&180 (dev-2)\\&&1.6&6.5&54&1040 (dev-3)\\
		\hline\hline
		SmB$_6$ \cite{Hong_KIST_arXiv_2018} & NiFe $|$ AlO$_x$ (0.38) & 0.004 & 500 & - &  200 \\ 
		\hline\hline
		Cu$|$Bi \cite{IssaCuBi2016} & NiFe (0.31) \cite{PhysRevB.87.094417} & $1\times10^{-4}$ & 0.15 & 100 & $1.5\times10^{-3}$ \\ 
		\hline
		Pt \cite{Pham_NanoLett_2016} & CoFe (0.58) & $8\times10^{-3}$ & 0.4 & 7 & 0.32 \\ 
		\hline
		Pt \cite{Liu_NPhys_2014} & CoFeB $|$ MgO (0.6) & $0.8\times10^{-3}$ & 8 & 7 & 0.64 \\ 
		\hline
		Ta \cite{Liu_NPhys_2014} & CoFeB $|$ MgO (0.6) & $6\times10^{-3}$ & 8 & 7 & 4.8 \\ 
		\hline
		W \cite{Liu_NPhys_2014} & CoFeB $|$ MgO (0.6) & $1.2\times10^{-3}$ & 8 & 7 & 0.96 \\ 
		\hline
		Ir \cite{Liu_NPhys_2014} & CoFeB $|$ MgO (0.6) & $0.6\times10^{-3}$ & 8 & 7 & 0.48 \\ 
		\hline\hline
		WTe$_2$ \cite{Li_NatComm_2018} & NiFe (0.45) & 0.14 & 5 & 23 & 70\\
		\hline
		Cd$_3$As$_2$ \cite{PhysRevLett.124.116802} & Co$|$oxide (0.4) & 40 & 0.15$^\dagger$ & - & 600\\
		\hline\hline
		InAs \cite{Koo_New} & NiFe $|$ Al$_2$O$_3$ (0.5) & 0.56 & 8 & 2 & 448 \\ 
		\hline
		
	\end{tabular}
\end{center}
\begin{flushleft}
        {\scriptsize $^\dagger$Diameter of the nanowire.}
    \end{flushleft}
\end{table*}

\section{Direct Effect: Charge to Spin Conversion}

\subsection{Figure-of-Merit for charge current to spin voltage conversion}

\subsubsection{Model}
A charge current $I_{12}$ in a spin-orbit (S-O) material induces a spin voltage $v_s$ \cite{Sayed_PRAppl_2018,Sayed_SciRep_2016}. In the absence of any adjacent layer, i.e., $G_L\rightarrow0$ in Fig. \ref{fig_circuit}(b), $v_s$ is given by
\begin{equation}
\label{spin_volt}
    v_s = \dfrac{\alpha p_{0,eff}}{2G_B}I_{12}.
\end{equation}
The spin voltage in the channel can be measured in the form of an open circuit charge voltage using an FM contact with its magnetization ($m$) along the spin polarization axis ($\hat{s}$) in the channel \cite{Hong_PRB_2012, Sayed_SciRep_2016}. The charge voltage difference between the two magnetic states ($+m$ and $-m$) is proportional to $v_s$, as given by (see Fig. \ref{spin_sig}(a))
\begin{equation}
    \label{Seokmin}
    \Delta V = V_{32}\left(+ m\right) - V_{32}\left(- m\right) = \dfrac{\alpha \xi p_{0,eff} p_f}{G_B}I_{12},
\end{equation}
where $\xi={G_{so}}/\left({G_{so} + G_L}\right)$. The method has been used on diverse classes of materials, including, topological insulators \cite{JonkerNatNano2014,KLWangNanoLett2014, DasNanoLett2015, SamarthPRB2015, YPChenSciRep2015, Samarth_PRB_2015, YoichiPRB2016}, transition metals \cite{Liu_NPhys_2014, Pham_APL_2016,Pham_NanoLett_2016}, semiconductors \cite{Koo_New}, Kondo insulators \cite{Hong_KIST_arXiv_2018}, and semimetals \cite{Li_NatComm_2018}, using both potentiometric and ohmic FM contacts. 

The figure-of-merit for a material's capability to convert a charge current into a spin voltage is $\Delta R_S = {\Delta V}/{I_{12}}$, which we define as the charge-spin interconversion resistance. We multiply both sides with $w$ to make it in the unit of resistivity
\begin{equation}
\label{spin_vol_FOM}
	\Delta R_S \times w = \dfrac{\Delta V}{I_{12}}\times w = \dfrac{h}{q^2}\dfrac{\alpha \xi p_{0,eff} p_f}{M_T/w},
\end{equation}
which also ensures that the experimental data points discussed in Fig. \ref{spin_sig}(b) are independent of the channel width. 

Eq. \eqref{Seokmin} describes a charge current induced spin voltage in a conductor with non-zero density of states ($D$) around the Fermi energy. For a fixed charge current ($I_{12}$), the spin voltage scales inversely with $M_T$ or $D$ around the Fermi energy and we expect a large spin voltage in a material with an arbitrarily small $D$.  However, a lower $D$ also yields a higher channel resistivity and a larger voltage drop across the channel length ($V_1-V_2$) is required to inject a sizable $I_{12}$ in the channel. For negligible spin current ($i_s\approx 0$), we can derive an expression for $\Delta V / \left(V_1-V_2\right)$ from Eq. \eqref{R_Mat} as
\begin{equation}
\label{ratio}
    {\left. {\frac{{\Delta V}}{{{V_1} - {V_2}}}} \right|_{{i_s} = 0}} = 2{p_f}\frac{{{\lambda _{IREE}}}}{{{\lambda _m} + L}},
\end{equation}
where $\lambda_{IREE}$ is given by Eq. \eqref{IEE_length} and will be discussed later. Note that the ratio of the open circuit spin voltage to the applied charge voltage across the channel length in Eq. \eqref{ratio} is independent of $M_T$.

%\scalebox{0.7}{
\begin{table*}
%		\captionsetup{justification=centering}
\addtolength{\tabcolsep}{-3pt}
\scriptsize
\caption{Estimation of Number of Modes and Degree of Spin-Momentum Locking in Diverse Materials.}\label{tab_Exp_Modes}
\begin{center}
	\begin{tabular}{||c | c | c |c |c|c| c| c| c||} 
		\hline
		\label{delRs_table_modes}
		SOC Material & $n$ & $n_s$ & $k_F$  $^i$ & $\rho$ & $R_{sheet}$ & $\lambda_m$ & $M_T/w$ & $p_{0,eff}'$ \\&($\times$10$^{25}$ m$^{-3}$)&($\times10^{16}$ m$^{-2}$)&(nm$^{-1}$)&($\mu\Omega$-cm)&($\Omega$)&(nm)&(nm$^{-1}$) &(Eq. \eqref{spin_vol_FOM})\\ [0.5ex] 
		\hline\hline
		(Bi$_{0.5}$Sb$_{0.5}$)$_2$Te$_3$ & - & - & - & - & 6000 \cite{SamarthPRB2015} & 150 \cite{SamarthPRB2015} & 0.029 (Eq. \eqref{Mt_rho}) & 0.56\\
		\hline
		(Bi$_{0.53}$Sb$_{0.47}$)$_2$Te$_3$ & 0.47$^b$ & - & 0.52 & - & - & - & 0.39 (Eq. \eqref{Mt_n}) & 0.73 \\ 
		\hline
		Bi$_{2}$Se$_{3}$ & 1.25 \cite{vanWees_PRB_2015} & - & 0.72 & - & - & - & 1.64 (Eq. \eqref{Mt_n}) & 0.066\\ 
		\hline
		Bi$_{2}$Se$_{3}$ & 5 \cite{DasNanoLett2015} & - & 1.14 & - & 18 \cite{DasNanoLett2015} & 130 \cite{SamarthPRB2015} & 8.3 (Eq. \eqref{Mt_n}) & 0.88\\ &&&&&&&11 (Eq. \eqref{Mt_rho})&\\
		\hline
		Bi$_{2}$Se$_{3}$ & 3 \cite{SamarthPRB2015} & - & 0.96 & - & - & - & 1 (Eq. \eqref{Mt_n}) & 0.12\\ 
		\hline
		Bi$_2$Te$_2$Se & - & - & 0.44 \cite{PhysRevB.82.241306} & - & - & - & 0.31 (Eq. \eqref{Mt_n}) & 0.83\\ 
		\hline
		BiSbTeSe$_2$ & 0.26$^c$ (dev-1)& - & 0.43 & - & - & - & 4.95 (Eq. \eqref{Mt_n}) & 0.073\\ & 0.1$^c$ (dev-2)&-&0.31&-&-&-&1.25 (Eq. \eqref{Mt_n}) & 0.3\\& 0.056$^c$ (dev-3)& - & 0.25 & - & - & - & 0.56 (Eq. \eqref{Mt_n}) & 0.79\\
		\hline\hline
		SmB$_6$ & - & - & 6.7 \cite{Hong_KIST_arXiv_2018} & - & - & - & 2.13 (Eq. \eqref{Mt_n}) & 0.68\\ 
		\hline\hline
		Cu$|$Bi & 8470$^d$ \cite{Ashcroft1976} & - & 13.6 & - & - & - & 2938 (Eq. \eqref{Mt_n}) & 0.0086$^e$\\ 
		\hline
		Pt & - & - & - & 27 \cite{Liu_NPhys_2014} & - & 10.5$^a$ \cite{Johann_PRB_1980} & 63.9 (Eq. \eqref{Mt_rho}) & 0.021 (from \cite{Pham_NanoLett_2016})\\&1600 \cite{Johann_PRB_1980}&-&7.8&-&-&-&67.7 (Eq. \eqref{Mt_n})&0.044 (from \cite{Liu_NPhys_2014})\\ 
		\hline
		Ta & - & - & - & 210 \cite{Liu_NPhys_2014} & - & 1.8$^f$ & 47.9 (Eq. \eqref{Mt_rho}) & 0.23\\
		\hline
		W & - & - & - & 42 \cite{Liu_NPhys_2014} & - & 2.1$^f$ & 205.4 (Eq. \eqref{Mt_rho}) & 0.2\\
		\hline
		Ir & - & - & - & 28 \cite{Liu_NPhys_2014} & - & 1.32$^g$ & 490.2 (Eq. \eqref{Mt_rho}) & 0.24\\ 
		\hline\hline
		WTe$_2$ & - & 1.9 \cite{Li_NatComm_2018} & 0.35 & - & - & - & 9.5 (Eq. \eqref{Mt_n}) & 0.9\\
		\hline
		Cd$_3$As$_2$ & 0.036 \cite{PhysRevLett.124.116802} & - & 0.22 & - & - & - & 1.02$^h$ (Eq. \eqref{Mt_n}) & 0.93\\
		\hline\hline
		InAs & - & 2 \cite{Koo_New} & 0.36 & - & - & - & 0.23 (Eq. \eqref{Mt_n}) & 0.12\\ 
		\hline
		
	\end{tabular}
	\begin{flushleft}
        {\scriptsize $^a$The reported mean free path is for samples with similar resistivity ($\sim20$ $\mu\Omega$-cm). $^b$Calculated using $n=n_s/t$ with $n_s=4.25\times10^{16}$ m$^{-2}$ \cite{KLWangNanoLett2014}. $^c$We have used $B/(ntq)$ to extract $n$ from their ordinary Hall resistance measurements on three different devices (dev-1, dev-2, and dev-3), where $B$ is the external magnetic field. $^d$ We assume that most of the current conduction occurs in the conductive layer Cu and calculated the total number of modes based on Cu parameters. $^e$ $p_0$ estimated from Eq. \eqref{RashbaSML} is $\sim0.296$ using $\alpha_R=3.2\times10^{-10}$ eV-m \cite{doi:10.1063/1.4828865} and $v_F=1.57\times10^6$ ms$^{-1}$ for Cu \cite{Ashcroft1976}. Note: $v_F=1.87\times10^6$ ms$^{-1}$ for Bi \cite{Ashcroft1976}. Estimation from the experiment in Ref. \cite{IssaCuBi2016} is much lower than this theoretical value and could be due to a higher current shunting in the magnetic contact. $^f$ We have assumed $\lambda_m$ to be equal to the spin diffusion lengths reported in Refs. \cite{Ralph_Science_2012} and \cite{Ralph_APL_2012}, it was pointed out previously that they are comparable \cite{PhysRevB.72.212410}. $^g$ $\lambda_m$ for Ir reported in Ref. \cite{Gall2016} is 7.09 nm for a sample with $\rho=$ 5.2 $\mu\Omega$-cm. We have assume a 5.4 times smaller value since the resistivity of the sample in Ref. \cite{Liu_NPhys_2014} is 5.4 times higher. $^h$ The channel is a nanowire with a cross-section $\pi r^2$ ($r$ is the radius). We have approximated $w=t=\sqrt{\pi}r$. $^i$ These values are estimated using $k_F = \sqrt[3]{3 \pi^2 n}$ or $k_F = \sqrt{2 \pi n_s}$ \cite{Sayed_PRAppl_2018}} or taken from the literature.
    \end{flushleft}
\end{center}
\end{table*}

\subsubsection{Comparison with experiments}
In Fig. \ref{spin_sig}(b), we compare Eq. \eqref{spin_vol_FOM} with experimentally measured $\Delta R_s\times w$ on a wide variety of materials as a function of $M_T/w$ of the corresponding device. We consider a diverse classes of materials with large range of variations in the density of states, including, topological insulators ((Bi$_{0.5}$Sb$_{0.5}$)$_2$Te$_3$ \cite{SamarthPRB2015}, (Bi$_{0.53}$Sb$_{0.47}$)$_2$Te$_3$ \cite{KLWangNanoLett2014}, Bi$_{2}$Se$_{3}$ \cite{vanWees_PRB_2015, DasNanoLett2015, SamarthPRB2015}, Bi$_2$Te$_2$Se \cite{YPChenSciRep2015}, and BiSbTeSe$_2$ \cite{Ando_arXiv_2016}), transition metals (Ta \cite{Liu_NPhys_2014}, Pt \cite{Pham_NanoLett_2016,Liu_NPhys_2014}, W \cite{Liu_NPhys_2014}, and Ir \cite{Liu_NPhys_2014}), metallic interfaces (Cu$|$Bi \cite{IssaCuBi2016}), narrow bandgap semiconductors (InAs \cite{Koo_JAppPhys_2012, Koo_New}), topological Kondo insulators (SmB$_6$ \cite{Hong_KIST_arXiv_2018}), and semimetals (WTe$_2$ \cite{Li_NatComm_2018}, Cd$_3$As$_2$ \cite{PhysRevLett.124.116802}). We estimate $M_T$ from the measured carrier concentration using Eq. \eqref{Mt_n} or from the measured resistivity / sheet resistance and known mean free path using Eq. \eqref{Mt_rho}. The details of the data points and related estimations are summarized in Tables \ref{delRs_table}-\ref{delRs_table_modes}, with detailed footnotes. 

Remarkably, the spin voltage in diverse classes of spin-orbit materials (both topological and non-topological) are scaling inversely proportional to the total number of modes in the material, as described by Eq. \eqref{Seokmin}. Although the data points are scattered in nature due to a variation of $t$, $p_{0,eff}$, $p_f$, and $\xi$ from sample to sample, we observe a seven orders of magnitude enhancement in the charge-spin interconversion resistance ($\Delta R_s$) due to a seven orders of magnitude lowering in the number of modes ($M_T$) over the diverse classes of materials shown in Fig. \ref{spin_sig}(b). This implies that the density of states or number of modes plays a dominant role in determining $\Delta R_s$. This observation is interesting and could be useful for spin-voltage based logic and memory applications, because the density of states can be externally controlled with gate-induced electronic fields on semiconductors and using strain on various oxides .

Note that some data points on topological materials in Fig. \ref{spin_sig}(b) correspond to low-temperature experiments, and the spin signals in these experiments decreases with the temperature. The temperature-dependent degradation is known to arise from the topological surface states formation conditions, positioning of the Fermi level in the topological bands, the coexistence of parallel channels, contact polarization degradation, etc., which are related to $\xi p_{0,eff} p_f$ in Eq. \eqref{Seokmin}. In Fig. \ref{spin_sig}(b), we take the maximum reported values of $\Delta R_s$ in various materials and plot them against their corresponding number of modes to illustrate how the density of states play a crucial role in the scaling of the spin voltage.

\subsubsection{Anatomy of Eq. \eqref{spin_volt}}
Eq. \eqref{spin_volt} can be decomposed into (i) the standard mesoscopic view for charge current and (ii) the conventional definition of spin-momentum locking. The mesoscopic view \cite{Datta_ETMS_1997} assigns two electrochemical potentials, $\mu^+$, and $\mu^-$, to electronic states with positive and negative group velocities, respectively. The charge current in the channel is related to $\mu^+$ and $\mu^-$ as
\begin{equation}
\label{mesoscopic}
    I_{12} = \dfrac{G_B}{q}\left(\mu^+ - \mu^-\right)= \dfrac{q}{h}M_T\left(\mu^+ - \mu^-\right),
\end{equation}
which indicates that $\mu^+ - \mu^-$ is proportional to $I_{12}$ and inversely proportional to $M_T$. Note that Eq. \eqref{mesoscopic} is valid for transport in the linear regime in any materials, with or without the SML.

A perfect SML means electronic states with the positive group velocity are all up spin-polarized; hence, $\mu^+$ represents an up spin potential, $\mu_{up}$. On the other hand, electronic states with the negative group velocity are all down spin-polarized; hence, $\mu^-$ represents a down spin potential, $\mu_{dn}$. In such an ideal scenario, we have
\begin{equation*}
    \mu^+ - \mu^- = \mu_{up} - \mu_{dn},
\end{equation*}
where spin voltage is $v_s = \left(\mu_{up} - \mu_{dn}\right)/2$. However, in real materials, the spin voltage is reduced by a lower strength of SML ($p_0$), spin-dependent scatterings in the channel ($\epsilon$), and angular distribution of the spin polarized states around the spin quantization axis ($\alpha$), as given by
\begin{equation}
\label{SML_view}
    \mu_{up}-\mu_{dn} = \alpha p_{0,eff} \left(\mu^+ - \mu^-\right).
\end{equation}
Combining Eqs. \eqref{mesoscopic} and \eqref{SML_view} yields Eq. \eqref{spin_volt}. Note that the observation in Fig. \ref{spin_sig}(b) that the spin signals scales inversely with $M_T$, originates from the mesoscopic view of charge current transport in Eq. \eqref{mesoscopic}. This feature is reflected on $v_s$ when spin polarization becomes locked to the momentum.

\begin{figure*}
	%\vspace*{1cm}
	\includegraphics[width=0.8 \textwidth]{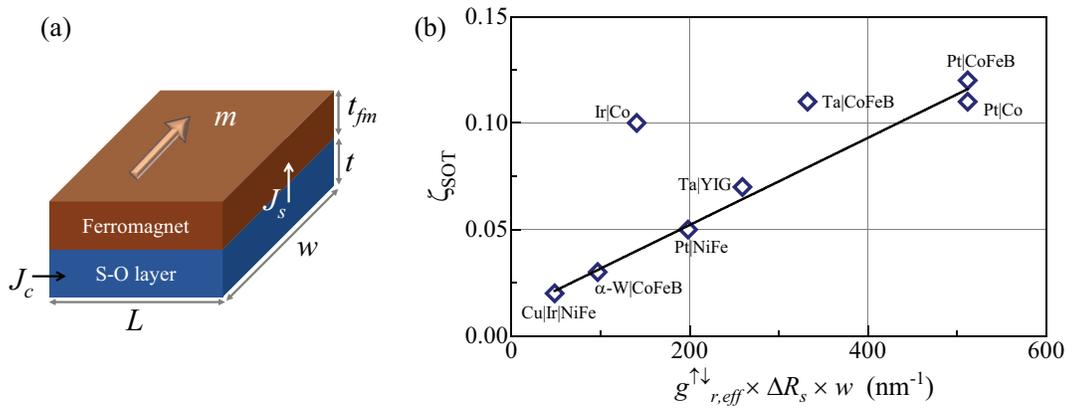}
	\centering
	\caption{(a) Structure for spin-orbit torque related experiments. (b) Spin-orbit torque (SOT) efficiency $\zeta_{SOT}$ in various metallic interfaces and comparison with  Eq. \eqref{SHA_spin}.}\label{fig_sha}
\end{figure*}

\subsection{Strength of the spin-momentum locking (SML) in various materials}

\subsubsection{SML strength in topological materials}
We have extracted the effective strength of SML $p_{0,eff}'=\xi p_{0,eff}$ from the experimental data points in Fig. \ref{spin_sig} using Eq. \eqref{spin_vol_FOM} and summarized them in Table  \ref{tab_Exp_Modes}. An ideal topological insulator is expected to show $p_{0,eff}=1$. The estimated values of $p_{0,eff}'$ from experiments are close to 1 for the materials known to exhibit topological surface states, e.g., (Bi$_{0.5}$Sb$_{0.5}$)$_2$Te$_3$, BiSbTeSe$_2$, Bi$_2$Te$_2$Se, and Bi$_{2}$Se$_{3}$. Although, $p_{0,eff}'$ estimated from the spin voltage measured on the Bi$_2$Se$_3$ sample in Ref. \cite{vanWees_PRB_2015} is much lower and could originate from a Rashba channel that can coexist in such material \cite{Bahramy2012, Sayed_NatComm_2019}. Note that the $p_{0,eff}'$ estimated on other topological materials are also close to unity, e.g., topological Kondo insulator SmB$_6$ \cite{Hong_KIST_arXiv_2018}, topological Weyl semimetal WTe$_2$ \cite{Li_NatComm_2018}, and a topological Dirac semimetal Cd$_3$As$_2$, indicating a strong SML as expected.

\subsubsection{SML strength in Rashba channels}
For the Cu$|$Bi interface, the estimated $p_{0,eff}'$ is $\sim$0.0086, which is very weak. Cu$|$Bi interface is a weak Rashba channel with a theoretical strength of the SML $p_0\approx {\alpha_R}/(\hbar{v_F}) \approx 0.054$ calculated using $\alpha_R=5.6\times10^{-11}$ eV-m \cite{IssaCuBi2016} and $v_F=1.57\times10^6$ ms$^{-1}$ (for Cu) \cite{Ashcroft1976} in Eq. \eqref{RashbaSML}. This lowering from the theoretical value can be attributed to a lower $\xi$ ($\sim$ 0.16) since the ferromagnet in Ref. \cite{IssaCuBi2016} was in direct contact with the metallic channel. Two-dimensional electron gas in InAs quantum-well is known as a strong Rashba channel, and $p_{0,\text{eff}}'$ estimated from the experiment is large $\sim$0.12.

\subsubsection{SML strength in metals}
The $p_{0,eff}'$ extracted for Pt is in the order of that observed in a weak Rashba channel. Measurements on Pt in Ref. \cite{Liu_NPhys_2014} were done with an oxide barrier at the interface and the extracted $p_{0,\text{eff}}'\approx 0.044$. However, the estimated $p_{0,\text{eff}}'\approx0.021$ is lower for Pt in Ref. \cite{Pham_NanoLett_2016}, where the ferromagnet was in direct contact with the channel. Such lowering could be attributed to a lower $\xi$ due to such direct contact. The extracted values of $p_{0,eff}'$ for Ta, W, and Ir are large ($\sim0.2$), in the order typically observed in a strong Rashba channel or a topological material with parallel channels. The origin of the SML in transition metals is a topic of active debate and could involve a bulk mechanism \cite{BauerPRB2013, Ralph_Science_2012} or an interface Rashba-like mechanism \cite{Felser_NatComm_2015, Grioni_PRL_2007, Saitoh_SciRep_2015, HoeschPRB2004, Tamai_PRB_2013}. Irrespective of the underlying mechanism, as long as there is a measurable spin voltage at the ferromagnetic contact, we can extract the strength of SML using Eq. \eqref{Seokmin}. 

\subsection{Figure-of-merit for the charge current to spin current conversion}

\subsubsection{Model for spin-orbit torque (SOT) efficiency}

The spin voltage, $v_s$, in the SO layer will inject a spin current, $i_s$, in the adjacent FM layer, given by
\begin{equation}
\label{direct_is}
    i_{s} = G_{s,eff} \times v_s.
\end{equation}
derived from the circuit in Fig. \ref{fig_circuit}(b). Here, $G_{s,eff}$ is given by Eq. \eqref{eff_source_cond}-\eqref{eff_source_cond_perp}  and $v_s$ is given by Eq. \eqref{spin_volt}-\eqref{Seokmin}. The spin current, $i_s$, applies a torque to the adjacent FM and the torque is maximum when $\vec m \perp \hat s$. A widely used figure-of-merit for charge current to spin current conversion is the SOT efficiency $\zeta_{SOT} = {J_s}/{J_c}$. We derive a model for $\zeta_{SOT}$ by combining Eq. \eqref{direct_is} with Eqs. \eqref{eff_source_cond_perp} and \eqref{spin_volt}, as given by
\begin{equation}
\label{SHA_spin}
    \left|\zeta_{SOT}\right| = \dfrac{\left|J_s\right|}{\left|J_c\right|} = \dfrac{q^2}{h} g_{r,eff}^{\uparrow\downarrow}\times \dfrac{\Delta R_s w t}{p_f},
\end{equation}
where $J_s = i_s / (w_fL_f)$ and $J_c = I_{12} / (w t)$. Interestingly, Eq. \eqref{SHA_spin} indicates $\zeta_{SOT}$ is determined by both the S-O layer and the magnetic interface and two independent measurements, $\Delta R_s$ from spin potentiometric experiments in Fig. \ref{spin_sig} and $g_{r,eff}^{\uparrow\downarrow}$ from FM resonance experiments, can be multiplied together to estimate the SOT efficiency in a device.  Eq. \eqref{SHA_spin} also indicates that it is possible that $\zeta_{SOT}$ is substantially different in a S-O material when coupled to a different FM interface.

We combine Eq. \eqref{SHA_spin} with Eq. \eqref{spin_vol_FOM} to get the following analytical expression, given by
\begin{equation}
\label{SHA}
    \left|\zeta_{SOT}\right| = \alpha p_{0,eff}' \times \dfrac{g_{r,eff}^{\uparrow\downarrow}}{m_n}.
\end{equation}
where $m_n=M_T/(wt)$ is the total number of modes per unit cross-sectional area of the channel. Note that $\zeta_{SOT}$ also scales inversely with the channel number of modes or the density of states around the Fermi energy in the material. This inverse scaling with $m_n$ can be related to the generally observed trend that resistive materials lead to higher SOT efficiency (see, e.g., Refs. \cite{PhysRevB.98.060410, Sinova_RMP_2015}), because the conductivity ($\sigma$) of the material is related to the $m_n$ as $\sigma = 1/\rho = \frac{q^2}{h}m_n \lambda_m$ (see Eq. \eqref{Mt_rho}).

We combine Eq. \eqref{SHA} with expression for $m_n$ in Eq. \eqref{Mt_n_3d} and get the following expression
\begin{equation}
\label{SHA_n}
    \left|\zeta_{SOT}\right| = \left(\dfrac{8}{3\pi^2}\right)^\frac{2}{3}\, \dfrac{p_{0,eff}' g_{r,eff}^{\uparrow\downarrow}}{n^\frac{2}{3}}.
\end{equation}
which indicates a scaling trend of $\zeta_{SOT}\propto n^{-\frac{2}{3}}$ with respect to the carrier concentration in the sample. Such a trend could be useful for future device design, as carrier concentration can be controlled with doping, electric gating, and strain modulation. Note that our model for the SOT efficiency describes the transport with electronic states only around the Fermi level, which is quite different from conventional views \cite{Sinova_RMP_2015, HoffmannIEETMAG2013} that considers all the occupied states. We will discuss on this difference in the subsection \ref{Bery}.

\subsubsection{SOT efficiencies in metals}

We have used $\Delta R_s$ data from Table \ref{tab_Exp_Stength} for S-O materials and calculated the effective SOT efficiency using reported $g_{r,eff}^{\uparrow\downarrow}$ on various S-O$|$FM interfaces. The calculations are summarized in Table \ref{tab_Exp_SHA}. The calculated values of $\zeta_{SOT}$ using Eq. \eqref{SHA_spin} have been compared with existing experiments on various metals in Table \ref{tab_Exp_SHA} and in Fig. \ref{fig_sha}, which show good agreement. Note that the calculated $\zeta_{SOT}$ for W is close to the experimental report for $\alpha$-W because the spin voltage value used was measured in Ref. \cite{Liu_NPhys_2014} on a sample with resistivity similar to that typically observed in $\alpha$-W \cite{ParkinAmorphous}. $\zeta_{SOT}$ is higher on resistive W samples, e.g., $\beta$-W \cite{Ralph_APL_2012} or amorphous $a$-W \cite{ParkinAmorphous}. In Fig. \ref{fig_sha}, we have compared with experiments on well-known metallic systems: Pt$|$NiFe, Pt$|$Fe-Co-B, Pt$|$Co, Ta$|$Co-Fe-B, Ta$|$YIG, W$|$Co-Fe-B, and Ir$|$Co, which show reasonably good agreement. Note that Eq. \eqref{SHA_spin} is applicable to other emerging materials  as well, which we have discussed in Table \ref{tab_Exp_SHA}-\ref{tab_SHA_Rashba}.

We have estimated $p_{0,eff}'$ for various known metallic Rashba interfaces from the reported Rashba coefficient $\alpha_R$ and Fermi velocity $v_F$ using Eq. \eqref{RashbaSML}. Interestingly, the calculated $\zeta_{SOT}$ using such theoretical estimations of $p_{0,eff}'$ matches reasonably well with measured $\zeta_{SOT}$ on Au$|$FM, and Ag$|$Bi$|$NiFe. Here, $m_n$ has been estimated from the carrier density of the corresponding conductive layers using Eq. \eqref{Mt_n}, and $g_{r,eff}^{\uparrow\downarrow}$ has been taken from measured values in the literature. The results are summarized in Table \ref{tab_Rashba_SHA}. $\zeta_{SOT}$ estimated from the Cu$|$Bi Rashba interface is of the same order as values observed in Au$|$FM, and Ag$|$Bi$|$NiFe. We note that the estimation of $\zeta_{SOT}$ for a Cu$|$Bi interface is $3$ times lower than the experimental observation on CuBi alloy, which indicates that such alloy has a different origin of SML, e.g., the resonant scattering from the Bi impurities, as discussed in Ref. \cite{PhysRevLett.109.156602}. Such scattering induced charge-spin interconversion is included in our general model and arises due to $p_{rs}$ as described in Section \ref{sec_SML} and Appendix \ref{AppA}. 

\subsubsection{SOT efficiencies in oxides}
\begin{figure}
	%\vspace*{1cm}
	\includegraphics[width=0.49 \textwidth]{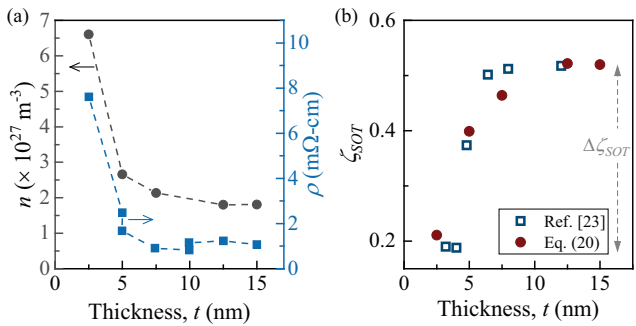}
	\centering
	\caption{(a) Resistivity and Hall carrier concentration of bare SrIrO$_3$ as a function of thickness, taken from Ref. \cite{Arnoud_PRMat_2019}. (b)  Spin-orbit torque (SOT) efficiency $\zeta_{SOT}$ in SrIrO$_3$ calculated using Eq. \eqref{SHA} with carrier concentration in Fig. \ref{fig_sio}(a), and comparison with experiments in Ref. \cite{Nan16186}.}\label{fig_sio}
\end{figure}

Recently, there is an increasing interest in transition metal oxides (see, e.g., \cite{Arnoud_PRMat_2019, Nan16186,PhysRevApplied.12.034004,FertLAOSTO2016}) for tunable charge-spin interconversion. In the semimetallic phase of strontium iridate (SrIrO$_3$), it has been observed that $\zeta_{SOT}$ increases for thicker devices while $\rho$ of the sample decreases \cite{Arnoud_PRMat_2019,Nan16186}, see Fig. \ref{fig_sio}(a). This observation is counter-intuitive to the observations in metals \cite{PhysRevB.98.060410} where higher resistivity exhibits higher SOT efficiency. Interestingly, the Hall carrier concentration measured on bare SrIrO$_3$ \cite{Arnoud_PRMat_2019} indicated that the concentration is also decreasing for thicker samples (see Fig. \ref{fig_sio}(a)). A similar observation that the resistivity and carrier concentration scale in the same direction with thickness has previously been observed in Bi$_2$Se$_3$ \cite{PhysRevB.84.073109} and Bi$_2$Te$_3$ \cite{doi:10.1002/adma.201501350}.

Eq. \eqref{SHA_n} indicates a scaling trend that $\zeta_{SOT}\propto n^{-\frac{2}{3}}$. We have calculated the SOT efficiency using Eq. \eqref{SHA_n} from the measured $n$ with varying sample thickness (see  Fig. \ref{fig_sio}(a)) and compared it with the SOT efficiency measurements on SrIrO$_3$ reported in Ref. \cite{Nan16186}, as shown in Fig. \ref{fig_sio}(b). We have set $p_{0,eff}'g_{r,eff}^{\uparrow\downarrow}\approx1.9\times10^{18}$ m$^{-2}$ to match the absolute value of $\zeta_{SOT}$ and assumed it to be constant over various thicknesses. The assumption that $g_{r,eff}^{\uparrow\downarrow}$ is constant over various thicknesses is based on the experimental observation in Ref. \cite{Nan16186}. It is surprising to note that two features observed in the SOT measurements were captured in the calculations using Eq. \eqref{SHA_n}.
\begin{enumerate}
    \item Independently measured $n$ decreases with thickness, and saturates for thicker samples (see Fig. \ref{fig_sio}(a)). Based on this observation, Eq. \eqref{SHA_n} indicates that the SOT efficiency will increase with the thickness and will saturate around the same thickness where $n$ saturates. Interestingly, the calculated $\zeta_{SOT}$ using Eq. \eqref{SHA_n} shows good agreement with the measurements reported in Ref. \cite{Nan16186}.

    \item The measured change in the SOT efficiency ($\Delta \zeta_{SOT}$ in Fig. \ref{fig_sio}(b)) between the thin and the thick limits is roughly the same as the value calculated using the change in $n$.
\end{enumerate}
The origin of the carrier concentration change with thickness needs a careful evaluation in the future and could arise from a phase change from tetragonal to orthorhombic \cite{Nan16186}, or a change in strain in the system \cite{Ramesh_strain_paper}. It has been discussed in the past \cite{Ramesh_strain_paper} that a strain on SrIrO$_3$ can significantly modulate the carrier concentration, i.e., the density of states in the material, which could be a promising way to achieve high $\zeta_{SOT}$ in an oxide system. Interestingly, such a strain induced tunability in the inverse voltage has been shown recently \cite{doi:10.1063/5.0027125}.

\begin{table*}
\caption{SOT efficiencies in various materials.}\label{tab_Exp_SHA}
\begin{center}
	\begin{tabular}{||c | c | c |c |c|c| c| c| c||} 
		\hline
		SOC Material & $\dfrac{\Delta R_s\times w\times t}{p_f}$ & $g_{r,eff}^{\uparrow\downarrow}$ &$\left|\zeta_{SOT}\right|$&$\left|\zeta_{SOT}\right|$\\&($\Omega$-nm$^{2}$)&(nm$^{-2}$)&(from  \eqref{SHA_spin}) & (measured) \\[0.5ex] 
		\hline\hline
		Pt $|$ NiFe & 74.7 & 15.2 \cite{Parkin_NPhys_2015} & 0.044 & 0.05 \cite{Parkin_NPhys_2015}\\ Pt $|$ FeCoB & & 40 \cite{doi:10.1063/1.4918909} & 0.115 & 0.12  \cite{doi:10.1063/1.4922084}\\
		Pt $|$ Co & & 39.6 \cite{Parkin_NPhys_2015}& 0.114 & 0.11 \cite{Parkin_NPhys_2015}  \\
		\hline
		Ta $|$ CoFeB & 560 & 6.92 \cite{Pandaeaav7200} & 0.15 & 0.12$\pm$0.04 \cite{Ralph_Science_2012}\\
		Ta $|$ YIG & & 5.4 \cite{WangPRL2014} & 0.11 & 0.07 \cite{WangPRL2014}\\ 
		\hline
		W $|$ CoFeB & 112 & 10.1 \cite{ParkSciRep2015} & 0.044 & 0.33 ($\beta$-W \cite{Ralph_APL_2012})\\
		 & & & & 0.03 ($\alpha$-W \cite{ParkinAmorphous})\\
		 & & & & 0.2$\sim$0.5 ($a$-W \cite{ParkinAmorphous})\\
		\hline
		Ir $|$ Co & 56 & 29.3 \cite{8399544} & 0.063 & 0.1 \cite{JimmySciRep2019}\\ 
		\hline
		Cu $|$ Ir $|$ NiFe & 56 & 10.1 \cite{PhysRevB.67.094421} & 0.022 & 0.02 \cite{PhysRevLett.106.126601}\\
		 \hline\hline
		 Bi$_{2}$Se$_{3}$ $|$ NiFe & 13333.33 &  12$\sim$65 \cite{Jamali2014}& 6.2$\sim$33.5 & 2$\sim$3.5 \cite{Mellnik2014a}\\
		  & 70000 & & 32.5$\sim$175.8 & 1.56$\sim$18.62 \cite{Mahendra_NatMat_2018}\\
		  & 13440 &  &6.2$\sim$33.8 &  \\
		\hline\hline
		WTe$_2$ & 35777.8 & 15.8 & 21.84 & 0.23$\sim$0.79 \cite{HYang_NatNano_2019}\\
		\hline
	\end{tabular}
\end{center}
\end{table*}
\begin{table*}
\footnotesize

\caption{SOT efficiencies in Rashba channels.}\label{tab_SHA_Rashba}
\begin{center}
	\begin{tabular}{||c | c | c |c |c|c| c|c| c| c| c||} 
		\hline
		\label{tab_Rashba_SHA}
		Material &$n$ & $k_F$ & $m_n$ $^{\ddag\ddag}$& $g_{r,eff}^{\uparrow\downarrow}$ & $\alpha_R$ & $v_F$ & $p_{0,eff}'$&$\left|\zeta_{SOT}\right|$&$\left|\zeta_{SOT}\right|$\\&($\times10^{28}$ m$^{-3}$)&(nm$^{-1}$)&(nm$^{-2}$)&(nm$^{-2}$)&(eV-pm)&($\times10^6$ ms$^{-1}$)&&(from Eq. \eqref{SHA}) & (measured) \\[0.5ex] 
		\hline\hline
		Au $|$ FM & 5.9$^\dagger$ \cite{Ashcroft1976} & 12 & 23 & 2.7 \cite{WangPRL2014} & 39.6 \cite{PhysRevB.88.125404} & 1.38 \cite{Gall2016} & 0.044 & 0.0033& 0.0033 \cite{Kwo_JAP_2013}\\& &&&&&&&& 0.0035 \cite{PhysRevB.82.214403,PhysRevB.88.064414}\\
		\hline
		Cu $|$ Bi $|$ NiFe & 8.47$^\dagger$ \cite{Ashcroft1976} & 13.6 & 29.4 & 10.1 \cite{PhysRevB.67.094421} & 320 \cite{doi:10.1063/1.4828865} & 1.57$^\dagger$ \cite{Ashcroft1976} & 0.296 & 0.07 & 0.24$^{\dagger\dagger\dagger}$  \cite{PhysRevLett.109.156602}\\ 
		\hline
		Ag $|$ Bi $|$ NiFe & 5.86$^\dagger$ \cite{Ashcroft1976} & 12 & 22.98 & 32.1 \cite{Fert_NatComm_2016} & 56 \cite{PhysRevB.88.125404} & 1.39$^\dagger$ \cite{Ashcroft1976} & 0.06 & 0.053 & 0.023 \cite{PhysRevB.89.054401}\\
		 \hline
		STO $|$ LAO $|$ NiFe & - & 1.28$^\ddag$ & 0.26 & 13.3 \cite{FertLAOSTO2016} & 3 \cite{FertLAOSTO2016} & 0.074$^{\dagger\dagger}$ & 0.062$^f$ & 2 & 1.8 \cite{PhysRevApplied.12.034004}\\
		\hline
	\end{tabular}
	\begin{flushleft}
        {\scriptsize$^\dagger$Parameters taken for the most conductive layer in the system. $^\ddag$Estimated using $k_F=\sqrt{2\pi n_s}$ with $n_s=2.6\times10^{17}$ m$^{-2}$ \cite{FertLAOSTO2016}. $^{\dagger\dagger}$Estimated using  $v_F=\hbar k_F /m^*$ with $m^*=2m_0$ \cite{FertLAOSTO2016}.$^{\ddag\ddag}$ Calculated using Eq. \eqref{Mt_n_3d}. $^{\dagger\dagger\dagger}$ This value was measured on a CuBi alloy, not on a Cu$|$Bi bilayer.}
    \end{flushleft}
\end{center}
\end{table*}

\subsubsection{Model parameters for $\zeta_{SOT}>1$}

In several of our calculations, we found $\zeta_{SOT}>1$ using parameters reported in the literature. For example, the $\Delta R_s$ for Bi$_2$Se$_3$ (in Table \ref{tab_Exp_Stength}) multiplied by the reported $g_{r,eff}^{\uparrow\downarrow}$ gives a calculated SOT efficiency $\zeta_{SOT}>1$, and similar high values have been observed experimentally (see, e.g., Refs. \cite{Mellnik2014a, Mahendra_NatMat_2018}). Also, the $\zeta_{SOT}$ estimated for LAO$|$STO Rashba interface using Eq. \eqref{SHA} is $>1$ and in agreement with the experimental observation in Ref. \cite{PhysRevApplied.12.034004}, see Table \ref{tab_Rashba_SHA}. Note that $p_{0,eff}\approx 0.062$ in LAO$|$STO and $\zeta_{SOT}>1$ arises due to $g_{r,eff}^{\uparrow\downarrow}\gg m_n$ in Eq. \eqref{SHA}. Also, the calculated $\zeta_{SOT}$ is $>1$ for WTe$_2$; however, the existing experimental report on WTe$_2$ is $<1$ (see, Ref. \cite{HYang_NatNano_2019}). We note that, for a proper estimation, the spin mixing conductance should be taken carefully since the reported values can be overestimated due to various non-ideal effects \cite{PhysRevLett.123.057203}. A more detailed analysis of Eq. \eqref{SHA} for the cases where $\zeta_{SOT}>1$ we leave for the future when more data are available as the field evolves.

\subsection{Comparison with conventional view}
\label{Bery}
\subsubsection{Effective SOT efficiency}
According to the semiclassical model in Ref. \cite{BauerPRB2013}, the SOT efficiency is given by \cite{doi:10.1063/1.4922084}
\begin{equation}
    \zeta_{SOT} = \theta_{SH} \frac{g_{r}^{\uparrow\downarrow}}{g_{r}^{\uparrow\downarrow}+g'_{so}} \tanh{\dfrac{t}{2\lambda_{sd}}}\tanh{\dfrac{t}{\lambda_{sd}}},
\end{equation}
where $\theta_{SH}$ is the internal spin Hall angle, $\lambda_{sd}$ is the spin-diffusion length, $g_{r}^{\uparrow\downarrow}$ is the real part of the bare spin-mixing conductance and $\frac{q^2}{h} g'_{so}=\frac{\sigma}{2 \lambda_{sd}} \tanh \left(\frac{t}{\lambda_{sd}}\right)$. In the thick S-O layer limit ($t\gg \lambda_{sd}$) and in the high resistivity S-O layer limit ($g'_{so}\ll g_{r}^{\uparrow\downarrow}$), we have
\begin{equation}
    \zeta_{SOT} \approx \theta_{SH}.
\end{equation}

Eq. \eqref{SHA} shows a similar dependence on $g_{r,eff}^{\uparrow\downarrow}=g_{r}^{\uparrow\downarrow}g_{so}/(g_{r}^{\uparrow\downarrow}+g_{so})$ where this term is determined by $\min(g_{r}^{\uparrow\downarrow},g_{so})$. In the limit where $g_{so}\ll g_{r}^{\uparrow\downarrow}$ and weak $p_{0,eff}'$,  reduces to
\begin{equation}
\label{SHA_red}
    \zeta_{SOT} \approx \alpha p_{0,eff}' \times \dfrac{g_{so}}{m_n},
\end{equation}
where $\frac{2q^2}{h}g_{so}L_fw_f=G_{so}$ is given by Eq. \eqref{source_cond}. Note that in this limit, Eq. \eqref{SHA_red} is determined completely by the S-O layer parameters.

Note that the limit $g_{so} \ll g_r^{\uparrow\downarrow}$ do not represent highly resistive channel within our model. When S-O layer have strong SML, even in the high resistivity limit we can satisfy $g_{so}\gg g_{r}^{\uparrow\downarrow}$ since $g_{so}\propto 1/(1-p_{0}^2)$, which yields
\begin{equation}
\label{SHA_red_new}
    \zeta_{SOT} \approx \alpha p_{0,eff}' \times \dfrac{g_{r}^{\uparrow\downarrow}}{m_n},
\end{equation}

\subsubsection{Internal spin Hall angle}

We can define an internal spin Hall angle in the weak SML limit from Eq. \eqref{SHA_red}, as
\begin{equation*}
    \theta_{SH} \equiv \alpha p_{0,eff}' \times \dfrac{g_{so}}{m_n}.
\end{equation*}

The internal spin Hall angle is often defined in terms of a spin Hall conductivity $\sigma_{SH}$ as
\begin{equation}
    \theta_{SH}=\dfrac{\sigma_{SH}}{\sigma}.
\end{equation}
Noting that $\sigma =\frac{q^2}{h} m_n \lambda_m$, we can also define a spin Hall conductivity from Eq. \eqref{SHA_red} as
\begin{equation}
\label{SHC}
    \sigma_{SH} = 2G'_{so}\lambda_{IREE}.
\end{equation}
where $G'_{so}=G_{so}/(w_f L_f)$. The model described in this paper considers transport near the Fermi energy and describes $\sigma_{SH}$ in terms of material density of states, degree of SML, mean free path, and spin source conductance. However, the conventional approach calculates $\sigma_{SH}$ using spin Berry phase from the electronic band structure and by taking into account contributions from anomalous velocities from all of the occupied states in the conduction band \cite{Sinova_RMP_2015, HoffmannIEETMAG2013}, including the states well below the Fermi energy. In the disordered phase of the material,  Bloch state description is not well-defined; however, the density of states and the number of modes are well-defined and measurable, even within a highly disordered sample.

\begin{table*}
\footnotesize
	\begin{center}
		\caption{Inverse Rashba-Edelstein effect (IREE) length in diverse materials.}
		\label{table_IREE}
		\begin{tabular}{||c | c | c | c | c ||} 
			\hline
			\label{IREE_table}
			Material & $\lambda_m$ (nm) & $p_{0,eff}'$ & $\lambda_{IEE}$ (nm) & $\lambda_{IEE}$ (nm)\\&&&(Eq. \eqref{IEE_length})&(measured)\\ [0.5ex] 
			\hline\hline
			Ag$|$Bi & 22.6  & 0.06 & 0.43 & 0.3 \cite{Fert_NatComm_2016}\\
			\hline
			Cu$|$Bi & 0.88 & 0.054 & 0.015 & 0.009 \cite{IssaCuBi2016}\\
			\hline
			Ag$|$Bi$_2$O$_3$ & 53.3  & 0.017 & 0.28 & 0.15$\pm$0.03 \cite{Tsai_SciRep_2018}\\
			\hline
			Cu$|$Bi$_2$O$_3$ & 39.9 & 0.024 & 0.3 & 0.17$\pm$0.03 \cite{Tsai_SciRep_2018}\\
			\hline
			Au$|$Bi$_2$O$_3$ & 37.7 & 0.011 & 0.13 & 0.09$\pm$0.03 \cite{Tsai_SciRep_2018}\\
			\hline
			Al$|$Bi$_2$O$_3$ & 18.9 & 0.004 & 0.024 & 0.01$\pm$0.002 \cite{Tsai_SciRep_2018}\\
			\hline
			Fe$|$Ge(111) & 3.27 & 0.11 &  0.12 & 0.13 \cite{Oyarzun_FeGe_2016}\\
			\hline
			MoS$_2|$Al & 40 & 0.3 & 3.82 & 4 \cite{Cheng_arXiv_2016} \\
			\hline
			LAO$|$STO & 180.4 & 0.062 & 3.56 & 6.4 \cite{FertLAOSTO2016}\\
			\hline
			Bi$_2$Se$_3$ & 6.26 & 0.066 & 0.13 & 0.035 \cite{SmarthPRL2016}\\
		     &  &  0.12 & 0.24 & 0.32 \cite{Mahendra_IEE_2019}\\
		     &  & 0.88 & 1.75 & \\
			\hline
		\end{tabular}
	\end{center}
	\begin{flushleft}
        {\scriptsize $\lambda_m$ and $p_{0,eff}'$ estimations are summarized in Table \ref{table_IEE_param}}
    \end{flushleft}
\end{table*}

\begin{table*}
	\begin{center}
	\scriptsize
	%\addtolength{\tabcolsep}{-3pt}
		\caption{Mean free path ($\lambda_m$) and effective degree of spin-momentum locking ($p_{0,eff}'$) in diverse materials.}
		\label{table_IEE_param}
		\begin{tabular}{||c | c | c |c | c |c | c |c | c | c |c|c|c | c ||} 
			\hline
			\label{IEE_table}
			SOC & $w$ & $t$ & $k_F$ & $G_B$ & $R$ & $R_{sheet}$ & $\rho$ & $\lambda_m$ & $\alpha_R$ & $v_F$ & $p_{0,eff}'$\\Material& ($\mu$m) & (nm) & (nm$^{-1}$)&&($\Omega$)&($\Omega/\square$)&($\mu\Omega$-cm)&(nm) &(eV$\cdot\AA$)&($\times10^6$ m$\cdot$s$^{-1}$)&\\ [0.5ex] 
			\hline\hline
			Ag$|$Bi & 400 \cite{Fert_NatComm_2016} & 5 \cite{Fert_NatComm_2016} & 12 & 1.77 kS & - & 10$^a$ & - & 22.6 & 0.56 \cite{Fert_NatComm_2016} & 1.39 \cite{Ashcroft1976} & 0.06\\
			\hline
			Cu$|$Bi & 0.15 \cite{IssaCuBi2016} & 20 \cite{IssaCuBi2016} & 13.6 &3.4 S& - & - & 100$^g$ & 0.88 & 0.56 \cite{IssaCuBi2016} & 1.57 \cite{Ashcroft1976} & 0.054\\
			\hline
			Ag$|$Bi$_2$O$_3$ & -& -& -&-&-&-&-& 53,3 \cite{Gall2016} & 0.16 \cite{Tsai_SciRep_2018} & 1.39 \cite{Ashcroft1976} & 0.017\\
			\hline
			Cu$|$Bi$_2$O$_3$ & -& -& -&-&-&-&-& 39.9 \cite{Gall2016} & 0.25 \cite{Tsai_SciRep_2018} & 1.57 \cite{Ashcroft1976} & 0.024\\
			\hline
			Au$|$Bi$_2$O$_3$ & -& -& -&-&-&-&-& 37.7 \cite{Gall2016}& 0.1 \cite{Tsai_SciRep_2018} & 1.4 \cite{Ashcroft1976} & 0.011\\
			\hline
			Al$|$Bi$_2$O$_3$ & -& -& -&-&-&-&-& 18.9 \cite{Gall2016}& 0.055 \cite{Tsai_SciRep_2018} & 2.03 \cite{Ashcroft1976} & 0.004\\
			\hline
			Fe$|$Ge(111) & 400 \cite{Oyarzun_FeGe_2016} & 20$^d$ \cite{Oyarzun_FeGe_2016} & 17.1$^e$ & 14.4 kS & 51 \cite{Oyarzun_FeGe_2016} & - & - &3.27$^f$ & 1.5$^b$ \cite{Lin_2014} & 1.98$^c$ \cite{Ashcroft1976} & 0.11 \\
			\hline
			MoS$_2|$Al &- &- &- & - &- &- & -& 40 \cite{Bhandari_2017} & 1.097 \cite{PhysRevB.91.125420} & 0.53 \cite{Ye_2017}& 0.3
			\\
			\hline
			LAO$|$STO & 400 \cite{FertLAOSTO2016} & - & 1.28 & 12.6 S & - & 176$^h$ & - & 180.4 & 0.03 \cite{FertLAOSTO2016} & 0.074$^i$ \cite{FertLAOSTO2016} & 0.062\\
			\hline
			Bi$_2$Se$_3$ & 1000 \cite{SmarthPRL2016} & 9 \cite{SmarthPRL2016} & 1.14 & 71.92 S & - & - & 2000$^j$ &  6.26 & - & - & 0.066$^k$\\
			 &  &  &  &  &  & &   &  & &  & 0.12$^k$\\
			 &  &  &  &  & &  &  &   &  &  & 0.88$^k$\\
			\hline
		\end{tabular}
	\end{center}
	\begin{flushleft}
		{\scriptsize$^a$Corresponds to sample with 5 nm Ag in Ref. \cite{Fert_NatComm_2016}.  $^b$Estimated for Ge interface with a metal which is higher than that reported for strained bulk Ge \cite{doi:10.1063/1.4901107}. $^c$ The Fermi velocity of the conductive layer Fe. $^d$Thickness of the most conducting layer is taken for calculation. $^e$Calculated using $k_F=\left(3\pi^2 n\right)^{\frac{1}{3}}$ from electron density of Fe: $n=1.7\times10^{29}$ m$^{-3}$ \cite{Ashcroft1976}. $^f$Estimated from $R=L/(G_B\lambda_m)$ with length $L=2.4$ mm \cite{Oyarzun_FeGe_2016}. $^g \rho$ of Bi layer was used which was taken from Ref. \cite{IssaCuBi2016}. $^h$Taken from Fig. 1(d) of Ref. \cite{FertLAOSTO2016} for LAO$|$STO at 7K. $^i$We estimate the Fermi velocity using ${{v}_{F}}=\frac{\hbar {{k}_{F}}}{{{m}^{*}}}$. $m^*\approx$ 2 $\times$ $9.1\times {{10}^{-31}}$ kg as reported in Ref. \cite{FertLAOSTO2016}. $^j$Taken from Fig. 2(b) of Ref. \cite{SmarthPRL2016} at $\sim$300K. $^k$Taken from Table \ref{tab_Exp_Modes}.}\\
	\end{flushleft}
\end{table*}

%\subsection{Comment on the reciprocity in direct and inverse effects}

%In the conventional view, the spin Hall effect (charge current induced spin current) and inverse spin Hall effect (spin current induced charge current) are related by the same parameter spin Hall angle ($\theta_{SH}$), given by
%\begin{subequations}
%\begin{equation}
%    J_s = \theta_{SH} J_c,
%\end{equation}
%\end{subequations}

\section{Inverse Effect: Spin to Charge Conversion}

\subsection{Spin Current to Charge Voltage}
A spin current $i_{s}$ injected into S-O layer from an FM layer will induce an open circuit charge voltage across the sample (see Fig. \ref{fig_circuit}(a)) as given by
\begin{equation}
\label{inv_volt}
    V_{12} = - \dfrac{\alpha \xi p_{0,eff}}{2 G_B} i_{s},
\end{equation}
where the spin current $i_s$ can be generated in various ways, e.g., spin pumping \cite{PhysRevLett.88.117601, RevModPhys.77.1375, Fert_NatComm_2016, FertLAOSTO2016}, spin Seebeck effect \cite{SaitohNature2008,Adachi_2013}, and electrical injection through a ferromagnetic contact \cite{IssaCuBi2016,Liu_NPhys_2014,Samarth_PRB_2015}. Eq. \eqref{inv_volt} satisfies the Onsager relation with Eq. \eqref{spin_volt} (see Ref. \cite{Sayed_SciRep_2016}).

\begin{figure*}
	%\vspace*{1cm}
	\includegraphics[width=0.8 \textwidth]{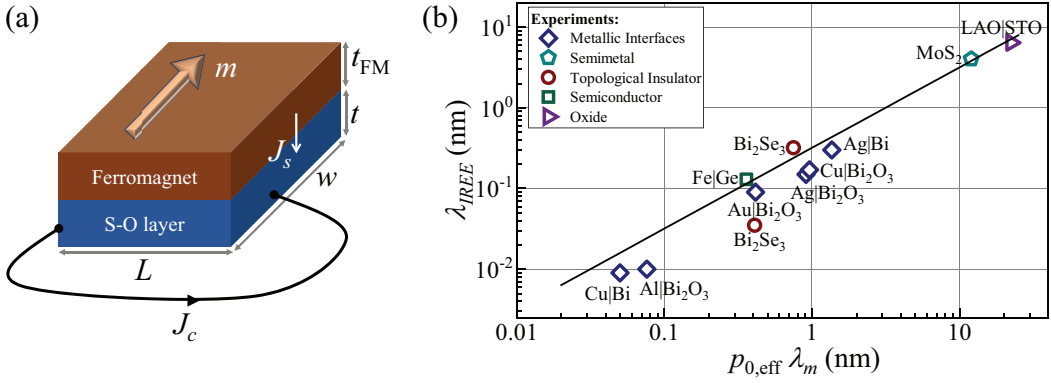}
	\centering
	\caption{(a) Structure for experiments on inverse effects. (b) Inverse Rashba-Edelstein effect (IREE) length in diverse classes of materials including Ag$|$Bi \cite{Fert_NatComm_2016}, Cu$|$Bi \cite{IssaCuBi2016}, Ag$|$Bi$_2$O$_3$ \cite{Tsai_SciRep_2018}, Cu$|$Bi$_2$O$_3$ \cite{Tsai_SciRep_2018},  Au$|$Bi$_2$O$_3$ \cite{Tsai_SciRep_2018},  Al$|$Bi$_2$O$_3$ \cite{Tsai_SciRep_2018}, LaAlO$_3|$SrTiO$_3$ (LAO$|$STO) \cite{FertLAOSTO2016}, Bi$_2$Se$_3$ \cite{SmarthPRL2016}, Fe/Ge(111) \cite{Oyarzun_FeGe_2016}, and MoS$_2$ \cite{Cheng_arXiv_2016, Mos2_2nd_paper}. The solid line represents Eq. \eqref{IEE_length}.}\label{fig_IREE}
\end{figure*}

The Onsager reciprocity \cite{Buttiker_PRB_2012, Sayed_SciRep_2016} requires that in Fig. \ref{fig_circuit}(a), we have
\begin{equation}
\label{Onsager}
    \dfrac{V_{32}(\vec m)}{I_{12}} = \dfrac{V_{12}(-\vec m)}{I_{32}},
\end{equation}
which in conjunction with Eq. \eqref{Seokmin} gives a reciprocal effect, given by
\begin{equation}
    \label{invSeokmin}
    V_{12}\left(+\vec m\right) - V_{12}\left(-\vec m\right) = - \dfrac{\alpha \xi p_{0,eff} p_f}{G_B}I_{32}.
\end{equation}
Eq. \eqref{invSeokmin} represents the inverse effect by electrical injection of a spin current $i_s \approx p_f I_{32}$ by flowing a current through the FM contact. Noting that $V_{12}\left(+\vec m\right) - V_{12}\left(-\vec m\right)\equiv 2V_{12}$, we get the Eq. \eqref{inv_volt}. The strength of the spin current induced charge voltage is the same as the charge current induced spin voltage and determined by the charge-spin interconversion resistance $\Delta R_s$ shown in Fig. \ref{spin_sig}(b), but differs by a negative sign. According to Eq. \eqref{Onsager}, we can write
\begin{equation*}
    \Delta R_s =\dfrac{V_{32}\left(+\vec m\right) - V_{32}\left(-\vec m\right)}{I_{12}}= -\dfrac{V_{12}\left(+\vec m\right) - V_{12}\left(-\vec m\right)}{I_{32}}.
\end{equation*}
Thus, the spin current induced charge voltage should show an inverse relation with the density of states of S-O material. It has been recently shown that the inverse spin Hall voltage in VO$_2$ exhibits an incremental jump while transitioning from metal to insulator phase \cite{Liu_NatComm2020}. This observation is in agreement with Eq. \eqref{inv_volt} since it is well known that the Hall carrier concentration in VO$_2$ shows a jump across the transition point \cite{PhysRevB.79.153107}.

\subsection{Spin Current to Charge Current}

The inverse effect in 2D channels with SML are often quantified with the following figure of merit
\begin{equation}
    \lambda_{IREE}=\dfrac{J_c}{J_s},
\end{equation}
known as the inverse Rashba-Edelstein effect (IREE) length. Here $J_c$ is the charge current density in the 2D channel (unit: A-m$^{-1}$) induced by the injected spin current density $J_s$ (unit: A-m$^{-2}$).

From Fig. \ref{fig_circuit}, if we connect the terminals 1 and 2, the short circuit charge current, $I_{sc}$, for a given spin current ($i_s$) injection is given by
\begin{equation}
    I_{sc} = \dfrac{G_B\lambda_m}{\lambda_m + L}\times \dfrac{\alpha p_{0,eff}'}{2G_B}i_s.
\end{equation}
For a diffusive channel ($L\gg \lambda_m$), we obtain an expression for $\lambda_{IREE}$, as given by
\begin{equation}
\label{IEE_length}
    \lambda_{IREE}=\dfrac{I_{sc}/w}{i_{s}/(wL)}=\dfrac{p_{0,eff}'\; \lambda_m}{\pi}.
\end{equation} 
Here, we compare Eq. \eqref{IEE_length} with available experiments on diverse classes of materials: Ag$|$Bi \cite{Fert_NatComm_2016}, Cu$|$Bi \cite{IssaCuBi2016}, Ag$|$Bi$_2$O$_3$ \cite{Tsai_SciRep_2018}, Cu$|$Bi$_2$O$_3$ \cite{Tsai_SciRep_2018},  Au$|$Bi$_2$O$_3$ \cite{Tsai_SciRep_2018},  Al$|$Bi$_2$O$_3$ \cite{Tsai_SciRep_2018}, LaAlO$_3|$SrTiO$_3$ (LAO$|$STO) \cite{FertLAOSTO2016}, Bi$_2$Se$_3$ \cite{SmarthPRL2016}, Fe/Ge(111) \cite{Oyarzun_FeGe_2016}, and MoS$_2$ \cite{Cheng_arXiv_2016, Mos2_2nd_paper}, which show good agreement as showin in Fig. \ref{fig_IREE}. The estimations are summarized in Table \ref{table_IREE}. Note that the figure-of-merit in Eq. \eqref{IEE_length} do not depend on the material density of states, but depend on the mean free path of the sample. One interesting observation in Table \ref{table_IREE} is that the $p_{0,eff}'$ in LAO$|$STO Rashba channel is weak. The large $\lambda_{IREE}$ observed in Ref. \cite{FertLAOSTO2016} is due to a large mean free path of the channel. %The delafossite metals can be a good candidate to explore higher $\lambda_{IREE}$ due to their high mean free path ($\sim$21 $\mu$m in PdCoO$_2$ and $\sim$4 $\mu$m PtCoO$_2$\cite{Mackenzie_2017}). Assuming $p_{0,eff}$ in PtCoO$_2$ is same as Pt, we expect $\lambda_{IREE}\approx 56 nm$.

\section{Summary}
The physics of charge-spin interconversion in various spin-orbit materials is a topic of great current interest for modern spintronics. Over the past decade, many materials have been studied, e.g., topological insulators, transition metals, Kondo insulators, semimetals, semiconductors, and oxides, to enhance interconversion efficiency. In this paper, we discuss a unified theoretical framework for such materials that relate the interconversion efficiency to the fundamental material parameters. We show that the charge-spin interconversion efficiency scales inversely with the channel number of modes or the material density of states near the Fermi energy. We further discuss two widely used figure of merits: spin-orbit torque (SOT) efficiency and inverse Rashba-Edelstein effect length for diverse classes of materials and how two enhance them in terms of materials and device parameters. Remarkably, experimental data obtained over the last decade on different materials closely follow our theoretical model which provides a unifying conceptual framework. We point out a scaling trend of the SOT efficiency with respect to the carrier concentration in agreement with the experiments. This unified model will enable a roadmap for materials with spin-orbit coupling and help design appropriate material systems and devices for desired spintronic applications.

\begin{acknowledgments}
	This work was supported by Applications and Systems-driven Center for Energy-Efficient integrated Nano Technologies (ASCENT), one of six centers in Joint University Microelectronics Program (JUMP), an Semiconductor Research Corporation (SRC) program sponsored by Defense Advanced Research Projects Agency (DARPA). S. Hong acknowledges support from the National Research and Development Program through the National Research Foundation of Korea (NRF), funded by the Ministry of Science and ICT (2019M3F3A1A02071509 and 2020M3F3A2A01081635) and KIST institutional program (2E31032). The authors are thankful to Professor Daniel C. Ralph in Cornell University, for the insightful discussions on the spin-orbit torque efficiencies in diverse classes of materials.
\end{acknowledgments}

\appendix

\section{Resistance Model and Circuit Representation}
\label{AppA}

\textit{In this section, we derive Eq. \eqref{R_Mat} from the semiclassical equations in Ref. \cite{Sayed_PRAppl_2018}.}

\subsection{Semiclassical Model}

\subsubsection{Diffusion equation for an SML channel}
\begin{figure}
	%\vspace*{1cm}
	\includegraphics[width=0.46 \textwidth]{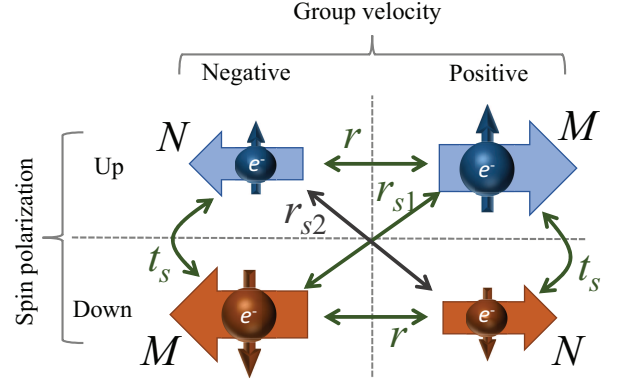}
	\centering
	\caption{Scattering mechanisms considered in the formalism among the four groups of electronic states, classified according to the spin polarization index (up or down) and the sign of the group velocity (positive or negative).}\label{fig_Scatter}
\end{figure}

We start from the diffusion equations for a general channel with spin-momentum locking \cite{Sayed_PRAppl_2018}, as given by
\begin{equation}
\label{semi_eq}
	\begin{aligned}
	&\frac{d}{{d x}}I_c = i^c,\\
	&\frac{d}{{dx}}{V_c} =  - \left( {\frac{1}{\lambda } + \frac{1}{{{\lambda _B}}}} \right)\frac{{{I_c}}}{{{G_B}}} - \frac{{\alpha {p_f}}}{{\lambda_B G_B}}{I_s} \\&+ \frac{{2}}{\alpha }\left( {\frac{1}{{{\lambda'}}} + \frac{{p_0}}{{{\lambda _B}}}} \right){V_s} + \frac{{2{p_0}{p_f}}}{{{\lambda _B}}}{V_c} + \frac{{\alpha {p_0}}}{{2{G_B}}}{i^s},\\
	&\frac{d}{{dx}}{I_s} =  - \frac{{4{G_B}}}{{{\alpha ^2}{\lambda _s}}}{V_s} + \frac{{2}}{{\alpha {\lambda _s'}}}{I_c} + {i^s},\\
	\text{and,}\;\;&\frac{d}{{dx}}{V_s} =  - \frac{{{\alpha ^2}}}{{{G_B}}}\left( {\frac{1}{{{\lambda _0}}} + \frac{1}{{{\lambda _B}}}} \right){I_s} - \frac{{\alpha {p_f}}}{{{\lambda _B}{G_B}}}{I_c} \\&+ 2\alpha {p_0}\left( {\frac{1}{{{\lambda _0}}} + \frac{1}{{{\lambda _B}}}} \right){V_c} + \frac{{2{p_0}{p_f}}}{{{\lambda _B}}}{V_s} + \frac{{\alpha {p_0}}}{{2{G_B}}}{i^c}.
	\end{aligned}
\end{equation}
Eq. \eqref{semi_eq} is obtained by combining Eqs. (8), (9), and (62) in Ref. \cite{Sayed_PRAppl_2018}, which were obtained from Boltzmann transport equation by classifying electronic states in the channel into four groups, based on their spin polarization index (up or down) and the sign of the group velocity (positive or negative), see Fig. \ref{fig_Scatter}. Here, $I_c, I_s$ are charge and spin currents in the channel, $V_c, V_s$ are charge and spin voltages in the channel, $G_B$ is given by Eq. \eqref{MT_Dv}, $p_0$ is given by Eq. \eqref{deg_SML}, and $\alpha$ is an angular averaging factor. $i^c, i^s$ are charge and spin current per unit length entering into the channel from an external contact with conductance per unit length $G_0$ and contact polarization $p_f$. The scattering length $\lambda_B$ in Eq. \eqref{semi_eq} is determined the conductance of the external contact with respect to the channel number of modes, as given by
\begin{equation}
    \dfrac{1}{\lambda_B} = \dfrac{G_0}{4G_B},
\end{equation}
where $G_0$ is the contact conductance per unit length of the contact. $\lambda$, $\lambda_0$, $\lambda_s$, $\lambda'$, and $\lambda_s'$ are scattering lengths in the channel which are described below.

\subsubsection{Scattering lengths in the channel}
We consider three types of scattering processes among the four groups of electronic states in the channel as shown in Fig. \ref{fig_Scatter}:
\begin{enumerate}
    \item transmission with spin-flip with scattering rate $t_s$,
    \item reflection without spin-flip with scattering rate $r$,
    \item and, reflection with spin-flip. Forward moving up spins become backward moving down spins and vice versa within the $M$ modes, with a scattering rate $r_{s1}$. Similarly, forward moving down spins become backward moving up spins and vice versa within the $N$ modes, with a scattering rate $r_{s2}$.
\end{enumerate}
Here, the scattering rates $r$, $t_s$, $r_{s1,2}$ are in the units of per unit length. The scattering lengths $\lambda$, $\lambda_0$, and $\lambda_s$ in Eq. \eqref{semi_eq} are given by \cite{Sayed_PRAppl_2018}
\begin{equation}
\label{mfps}
\begin{aligned}
&\dfrac{1}{\lambda } = \dfrac{1}{2}\left( {\dfrac{{{r_{s2}}}}{N} + \dfrac{{{r_{s1}}}}{M}} \right) + \dfrac{r}{2}\left( {\dfrac{1}{N}\; + \dfrac{1}{M}} \right),\\
&\dfrac{1}{{{\lambda _0}}} = \dfrac{r + {t_s}}{2} \left( {\dfrac{1}{N} + \dfrac{1}{M}} \right), \;\text{and}\\
&\dfrac{1}{{{\lambda _s}}} = \dfrac{1}{2}\left( {\dfrac{{{r_{s2}}}}{N} + \dfrac{{{r_{s1}}}}{M}} \right) + \dfrac{{t_s}}{2}\left( {\dfrac{1}{N} + \dfrac{1}{M}} \right).\,
\end{aligned}
\end{equation}
Here, $\lambda$ is the back scattering length for the charge transport, $\lambda_0$ is a scattering length for the spin transport, and $\lambda_s$ is a scattering length for spin-relaxation in the channel. The scattering lengths $\lambda'$ and $\lambda_s'$ in Eq. \eqref{semi_eq} are related to spin induced charge and charge induced spin, respectively. They are given by \cite{Sayed_PRAppl_2018}
\begin{equation}
\label{spin_charge_coupling}
\begin{array}{l}
\dfrac{1}{{\lambda '}} = \dfrac{1}{2}\left( {\dfrac{{{r_{s2}}}}{N} - \dfrac{{{r_{s1}}}}{M}} \right) + \dfrac{r}{2}\left( {\dfrac{1}{N}\; - \dfrac{1}{M}} \right),\;\text{and}\\
\dfrac{1}{{{{\lambda_s '}}}} = \dfrac{1}{2}\left( {\dfrac{{{r_{s2}}}}{N} - \dfrac{{{r_{s1}}}}{M}} \right) + \dfrac{t_s}{2}\left( {\dfrac{1}{N} - \dfrac{1}{M}} \right).
\end{array}
\end{equation}

For simplicity of the analytical details, we define
\begin{equation}
\label{def1}
\begin{array}{l}
\dfrac{1}{{\lambda_r}} = \dfrac{r}{2}\left( {\dfrac{1}{N} + \dfrac{1}{M}} \right),\\
\dfrac{1}{{{{\lambda_t}}}} = \dfrac{t_s}{2}\left( {\dfrac{1}{N} + \dfrac{1}{M}} \right),\;\text{and}\\ \dfrac{1}{{\lambda_{rs}}}=\dfrac{1}{2}\left( {\dfrac{{{r_{s2}}}}{N} - \dfrac{{{r_{s1}}}}{M}} \right).
\end{array}
\end{equation}
which allows us to re-write Eq. \eqref{spin_charge_coupling} as
\begin{equation}
\label{spin_charge_coupling1}
\begin{array}{l}
\dfrac{1}{{\lambda '}} = \dfrac{1}{{\lambda_{rs}}} + \dfrac{p_0}{{\lambda_r}},\;\text{and}\\
\dfrac{1}{{{{\lambda_s '}}}} = \dfrac{1}{{\lambda_{rs}}} + \dfrac{p_0}{{\lambda_t}}.
\end{array}
\end{equation}

Note that the resistance matrix similar to Eq. \eqref{R_Mat} presented in Ref. \cite{Sayed_PRAppl_2018} considered potentiometric (non-invasive) external contacts only, by assuming $\lambda_B\rightarrow\infty$. Also, to derive the resistance matrix, Ref. \cite{Sayed_PRAppl_2018} assumed that the spin-flip reflection is the dominant scattering, i.e. $\frac{1}{\lambda},\frac{1}{\lambda_s}\gg \frac{1}{\lambda_0},\frac{1}{\lambda'}, \frac{1}{\lambda_s'}$. In this manuscript, we present a resistance matrix applicable for general scattering conditions and contacts with arbitrary conductance.

%\begin{enumerate}
    %\item When the total spin-flip reflection rate within the forward moving up polarized states (or backward moving down polarized states) is same as the total spin-flip reflection rate within the backward moving up polarized states (or forward moving down polarized states), then 
    %\begin{equation}
    %%   \dfrac{1}{\lambda_r} = \dfrac{1}{\lambda}\;\;\; \text{and,}\;\;\; \dfrac{1}{\lambda_t} = \dfrac{1}{\lambda_s}.
    %\end{equation}
    
    %\item When the spin-flip reflection rate per mode within the forward moving up polarized states (or backward moving down polarized states) is same as the spin-flip reflection rate per mode within the backward moving up polarized states (or forward moving down polarized states), then
    %\begin{equation}
    %\label{assumption_2}
     %   \dfrac{1}{\lambda_r} =\dfrac{1}{2}\left( \dfrac{1}{\lambda}-\dfrac{1}{\lambda_s}+\dfrac{1}{\lambda_0}\right)\;\;\; \text{and,}\;\;\; \dfrac{1}{\lambda_t} =\dfrac{1}{2}\left( \dfrac{1}{\lambda_s}-\dfrac{1}{\lambda}+\dfrac{1}{\lambda_0}\right).
    %\end{equation}
%\end{enumerate}

\subsection{Assumptions}
We make the following two assumptions:
\begin{itemize}
    \item No charge current is flowing out of the external contact, i.e., $i^c=0$. The external contact can only inject or absorb a spin current $i^s$. 
    \item We consider a channel region where the spin voltage is uniform, i.e.
    \begin{equation*}
        \dfrac{d}{dx}V_s = 0.
    \end{equation*}
\end{itemize}
Thus the diffusion equations can be re-written as:
\begin{widetext}
\begin{subequations}
	\begin{alignat}{4}
	\label{Diff1}
	&\dfrac{d}{{d x}}I_c = 0,\\
	\label{Diff2}
	&\frac{d}{{dx}}{V_c} =  - \left( {\frac{1}{\lambda } + \frac{1}{{{\lambda _B}}}} \right)\frac{{{I_c}}}{{{G_B}}} - \frac{{\alpha {p_f}}}{{G_B\lambda_B}}{I_s} +\frac{2}{\alpha\lambda_{rs}} V_s + \frac{{2{p_0}}}{\alpha }\left( {\frac{1}{{{\lambda _r}}} + \frac{1}{{{\lambda _B}}}} \right){V_s} + \frac{{2{p_0}{p_f}}}{{{\lambda _B}}}{V_c} + \frac{{\alpha {p_0}}}{{2{G_B}}}{i^s},\\
	\label{Diff3}
	&\dfrac{d}{{dx}}{I_s} =  - \dfrac{{4{G_B}}}{{{\alpha ^2}{\lambda _s}}}{V_s} +\dfrac{2}{\alpha \lambda_{rs}}I_c + \dfrac{{2{p_0}}}{{\alpha {\lambda _t}}}{I_c} + {i^s},\\
	\label{Diff4}
	\text{and}\;\;&\frac{d}{{dx}}{V_s} = 0 =  - \frac{{{\alpha ^2}}}{{{G_B}}}\left( {\frac{1}{{{\lambda _0}}} + \frac{1}{{{\lambda _B}}}} \right){I_s} - \frac{{\alpha {p_f}}}{{{\lambda _B}{G_B}}}{I_c} + 2\alpha {p_0}\left( {\frac{1}{{{\lambda _0}}} + \frac{1}{{{\lambda _B}}}} \right){V_c} + \frac{{2{p_0}{p_f}}}{{{\lambda _B}}}{V_s}.
	\end{alignat}
\end{subequations}
\end{widetext}
Eq. \eqref{Diff4} can be simplified as:
\begin{equation}
    \label{Vs_eqn}
    {I_s} = \frac{{2{p_0}{G_B}}}{\alpha }{V_c} - \frac{{{\kappa p_f}}}{\alpha }{I_c} + \frac{{{\kappa p_f}}}{\alpha }\frac{{2{p_0}{G_B}}}{\alpha }{V_s}.
\end{equation}
where
\begin{equation}
    \kappa = \frac{{{\lambda _0}}}{{{\lambda _0} + {\lambda _B}}}.
\end{equation}

\subsection{Second row of the resistance matrix}
We differentiate both sides of Eq. \eqref{Vs_eqn} with respect to $x$ to get 
\begin{equation}
    \label{Vs_eqn_diff}
    \dfrac{d}{{dx}}{I_s} = \dfrac{{2{p_0}{G_B}}}{\alpha }\dfrac{d}{{dx}}{V_c},
\end{equation}
which we combine with Eqs. \eqref{Diff2} and \eqref{Diff3} to get
\begin{equation}
\label{relation1}
\begin{aligned}
    - \frac{{4{G_B}}}{{{\alpha ^2}}}\left[ {\frac{1}{{{\lambda _s}}} + {p_0}\left( {\frac{1}{{{\lambda _{rs}}}} + \frac{{{p_0}}}{{{\lambda _r}}} + \frac{{{p_0}}}{{{\lambda _B}}}} \right)} \right]{V_s} =  - \frac{{2{p_0}{p_f}}}{{{\lambda _B}}}{I_s} \\- \frac{2}{\alpha }\left( {\frac{1}{{{\lambda _{rs}}}} + {p_0}\left( {\frac{1}{{{\lambda _t}}} + \frac{1}{\lambda } + \frac{1}{{{\lambda _B}}}} \right)} \right){I_c} \\+ \frac{{4p_0^2{p_f}{G_B}}}{{\alpha {\lambda _B}}}{V_c} - \left( {1 - p_0^2} \right){i^s},
\end{aligned}
\end{equation}

We then combine Eqs. \eqref{relation1} and \eqref{Vs_eqn} to get
\begin{equation}
\label{rwo2}
\begin{aligned}
{V_s} =\frac{{\alpha {\lambda _{s0}}}}{{2{G_B}{\lambda _{rs}}}}{I_c} +  \frac{{\alpha\epsilon {p_0}}}{{2{G_B}}}{I_c} + \frac{{{\alpha ^2}\left( {1 - p_0^2} \right)\lambda_{s0}}}{{4{G_B}L}}{i_s},
\end{aligned}
\end{equation}
where $i_s = Li^s$, $\lambda_{s0}$ and $\epsilon$ are given by
\begin{subequations}
    \begin{equation}
    \dfrac{1}{\lambda_{s0}} ={\frac{1}{{{\lambda _s}}} + \frac{{{p_0^2}}}{{{\lambda _r}}} + \frac{{p_0^2}}{{{\lambda _B}}}\left( {1 - \kappa p_f^2} \right)},\;\;\; \text{and}
\end{equation}
\begin{equation}
    \label{xi}
    \epsilon= \frac{{{\lambda _{s0}}}}{{{\lambda _t}}} + \frac{{{\lambda _{s0}}}}{\lambda } + \frac{{{\lambda _{s0}}}}{{{\lambda _B}}}\left( {1 - \kappa p_f^2} \right).
\end{equation}
\end{subequations}

\subsection{First row of the resistance matrix}

We combine Eq. \eqref{Diff2} with Eq. \eqref{Vs_eqn} to get
\begin{equation}
\label{rel3}
\begin{aligned}
    \frac{d}{{dx}}{V_c} =  - \left( {\frac{1}{\lambda } + \frac{{1 - \kappa p_f^2}}{{{\lambda _B}}}} \right)\frac{{{I_c}}}{{{G_B}}} + \frac{2}{{\alpha {\lambda _{rs}}}}{V_s}\\ + \frac{{2{p_0}}}{\alpha }\left( {\frac{1}{{{\lambda _r}}} + \frac{{1 - \kappa p_f^2}}{{{\lambda _B}}}} \right){V_s} + \frac{{\alpha {p_0}}}{{2{G_B}}}{i^s}.
\end{aligned}
\end{equation}

We apply $\dfrac{d}{dx}V_c=-\dfrac{V_1-V_2}{L}$ and combine Eq. \eqref{rel3} with Eq. \eqref{rwo2} to get
\begin{equation}
\label{rwo1}
\begin{aligned}
{V_1} - {V_2} = \frac{L}{{{G_B}{\lambda _m}}}{I_c} - \frac{{\alpha {p_0}}}{{2{G_B}}}\left\{ {\frac{{{\lambda _{s0}}}}{{{\lambda _r}}} + \frac{{{\lambda _{s0}}}}{{{\lambda _s}}} + \frac{{{\lambda _{s0}}}}{{{\lambda _B}}}\left( {1 - \kappa p_f^2} \right)} \right\}{i_s} \\- \frac{{\alpha {\lambda _{s0}}}}{{2{G_B}{\lambda _{rs}}}}{i_s},
\end{aligned}
\end{equation}
where $i_{s}=Li^s$ and the effective mean free path ($\lambda_m$) is given by
\begin{equation}
\begin{aligned}
    \frac{1}{{{\lambda _m}}}& = \frac{1}{\lambda } - \frac{{p_0^2}}{{{\lambda _r}}} - \frac{{{\lambda _{s0}}}}{{\lambda _{rs}^2}} - \frac{{{p_0}}}{{{\lambda _{rs}}}} - \frac{{{p_0}{\lambda _{s0}}}}{{{\lambda _r}{\lambda _{rs}}}} \\&+ \frac{{1 - \kappa p_f^2}}{{{\lambda _B}}}\left( {\frac{{{\lambda _{s0}}}}{{{\lambda _s}}} + \frac{{{\lambda _{s0}}p_0^2}}{{{\lambda _r}}} - \frac{{{\lambda _{s0}}p_0^2}}{{{\lambda _t}}} - \frac{{{\lambda _{s0}}p_0^2}}{\lambda }} \right).
\end{aligned}
\end{equation}
From Eqs. \eqref{mfps} and \eqref{spin_charge_coupling} we see that
\begin{equation}
    \dfrac{1}{\lambda}+\dfrac{1}{\lambda_t}=\dfrac{1}{\lambda_s}+\dfrac{1}{\lambda_r},
\end{equation}
which allows us to write Eq. \eqref{rwo1} as
\begin{equation}
\label{rwo2a}
   {V_1} - {V_2} = \frac{L}{{{G_B}{\lambda _m}}}{I_c} - \frac{{\alpha \epsilon {p_0}}}{{2{G_B}}}{i_s} - \frac{{\alpha {\lambda _{s0}}}}{{2{G_B}{\lambda _{rs}}}}{i_s}.
\end{equation}

The terms related to $\lambda_{rs}$ in Eqs. \eqref{rwo2a} and \eqref{rwo2} indicate an additional component on charge-spin interconversion induced by spin-flip scatterings. We define the strength of such scattering induced charge-spin interconversion as
\begin{equation}
    p_{rs} = \dfrac{\lambda_{s0}}{\lambda_{rs}}.
\end{equation}
Thus we can write Eqs. \eqref{rwo2a} and \eqref{rwo2} as
\begin{subequations}
    \begin{equation*}
        {V_1} - {V_2} = \frac{L}{{{G_B}{\lambda _m}}}{I_c} - \frac{{\alpha{p_{0,eff}}}}{{2{G_B}}}{i_s},
    \end{equation*}
    \begin{equation*}
        {V_s} = \frac{{\alpha {p_{0,eff}}}}{{2{G_B}}}{I_c} + \frac{{{\alpha ^2}\left( {1 - p_0^2} \right)\lambda_{s0}}}{{4{G_B}L}}{i_s},
    \end{equation*}
\end{subequations}
which yields the resistance matrix in Eq. \eqref{R_Mat}, where $p_{0,eff}$ is given by
\begin{equation}
    p_{0,eff} = \epsilon p_0 + p_{rs},
\end{equation}
which indicate that the charge-spin interconversion can have a contribution from the band structure of the material (given by $p_0$) and also a spin-dependent scattering induced component (given by $p_{rs}$).

\subsection{Pure scattering induced charge-spin interconversion}
Eqs. \eqref{rwo2a} and \eqref{rwo2} indicate a pure scattering induced charge-spin interconversion, even in a normal metal (i.e., $p_0=0$ or $M=N$). Such charge-spin interconversion can be induced in a normal metal due to a difference in $r_{s1}$ and $r_{s2}$, i.e., forward (or backward) moving up-spins scatter at a different rate than the down-spins.. In a normal metal, Eqs. \eqref{rwo2a} and \eqref{rwo2} become
\begin{subequations}
\begin{equation}
\label{rwo2scat}
{V_1} - {V_2} = \frac{{ L}}{{{G_B}{\lambda _m'}}}{I_c}  - \frac{{\alpha p_{rs} }}{{2{G_B}}}{i_s}.
\end{equation}
\begin{equation}
\label{rwo1scat}
{V_s} =\frac{{\alpha p_{rs}}}{{2{G_B}}}{I_c} + \frac{{{\alpha ^2}\lambda_{s}}}{{4{G_B}L}}{i_s},
\end{equation}
\end{subequations}
where $\frac{1}{{{\lambda _m'}}} = \frac{1}{\lambda } + \frac{{{\lambda _s}}}{{\lambda _{rs}^2}}$, and $p_{rs}$ represents the strength of the spin-flip scattering induced charge-spin interconversion in a normal metal channel, given by
\begin{equation}
    p_{rs} = \dfrac{\left(r_{s2}+ts\right)-\left(r_{s1}+ts\right)}{\left(r_{s2}+ts\right)+\left(r_{s1}+ts\right)}.
\end{equation}
Eqs. \eqref{rwo2scat}-\eqref{rwo1scat} indicate a charge-to-spin and a spin-to-charge conversion in a normal metal channel, similar to Eq. \eqref{R_Mat} describing an SML channel.

The phenomena described by Eqs. \eqref{rwo2scat}-\eqref{rwo1scat} are aligned with experimental observations on impurity scattering induced high SOT (e.g., Ref. \cite{PhysRevLett.109.156602}), where a copper sample (a well-known normal metal channel) exhibits a large SOT by introducing bismuth impurities. Note that in the present manuscript, we have extracted $p_{0,eff}'$ from measurements and do not assume any particular origin.

\section{Number of Modes and Density of States}
\label{AppC}

The density of states in a material is given by
\begin{equation}
\label{DOS_T}
    D_0 = \dfrac{dN}{dE} = \dfrac{dN}{dp}\dfrac{dp}{dE},
\end{equation}
where $N$ is the number of electronic states in the channel, $p$ is the momentum, and $E$ is the energy. Noting that the group velocity $v=dE/dp$, Eq. \eqref{DOS_T} gives
\begin{equation}
\label{DOS_Tp}
    D_0\,v = \dfrac{dN}{dp}.
\end{equation}

The total number of states, $N(p)$, that have a momentum less than $p$ can be counted using the following formula for a 2D channel with length $L$ and width $w$ \cite{Datta_LNE_2012}
\begin{equation}
\label{Np_2d}
    N(p)=2\times\dfrac{\pi p^2}{\frac{h}{L}\frac{h}{w}},
\end{equation}
and using the following formula for a 3D channel with length $L$, width $w$, and thickness $t$
\begin{equation}
\label{Np_3d}
    N(p)=2\times\dfrac{\frac{4}{3}\pi p^3}{\frac{h}{L}\frac{h}{w}\frac{h}{t}},
\end{equation}
which in conjunction with Eq. \eqref{DOS_Tp} gives
\begin{subequations}
	\begin{equation}
	\label{Dv_2d}
		D\,v = 2\times\dfrac{2\pi p}{h^2} w,\;\;\;\;\;\;\text{(for a 2D channel)}
	\end{equation}
	\begin{equation}
	\label{Dv_3d}
		D\,v = 2\times\dfrac{4\pi p^2}{h^3} wt,\;\;\;\;\;\;\text{(for a 3D channel)}
	\end{equation}
\end{subequations}
where $D=D_0/L$ represents the density of states per unit length. Here, the pre-factor $2$ were introduced for two types of spins. 

Combining Eqs. \eqref{Dv_2d}-\eqref{Dv_3d} with Eq. \eqref{MT_Dv}
which in conjunction with Eq. \eqref{DOS_Tp} gives
\begin{subequations}
	\begin{equation}
	\label{MTp_2d}
		M_T = 2\times\dfrac{p}{\pi\hbar}w,\;\;\;\;\;\;\text{(for a 2D channel)}
	\end{equation}
	\begin{equation}
	\label{MTp_3d}
		M_T = 2\times\dfrac{p^2}{4\pi\hbar^2} wt,\;\;\;\;\;\;\text{(for a 3D channel)}
	\end{equation}
\end{subequations}
At the Fermi energy, we can write $p=\hbar k_F$ in terms of the Fermi wavevector $k_F$. $k_F$ is related to the carrier concentration in the channel as \cite{Sayed_PRAppl_2018}
\begin{subequations}
	\begin{equation}
	\label{kF_2d}
		k_F = \sqrt{2\pi n_s},\;\;\;\;\;\;\text{(for a 2D channel)}
	\end{equation}
	\begin{equation}
	\label{kF_3d}
	k_F = \sqrt[3]{3\pi^2 n},\;\;\;\;\;\;\text{(for a 3D channel)}
	\end{equation}
\end{subequations}
where $n_s$ and $n$ are the carrier concentrations for a 2D and a 3D channel. Combining Eqs. \eqref{kF_2d}-\eqref{kF_3d} to Eqs. \eqref{MTp_2d}-\eqref{MTp_3d} yields the expressions in Eqs. \eqref{Mt_n}.

\section{Spin-Momentum Locking in Rashba Channels}
\label{AppB}
We start from the following Rashba Hamiltonian
\begin{equation}
\label{RashbaH}
\mathcal{H}=\dfrac{\hbar^2k^2}{2m^*}I_{2\times2}-\alpha_R \left(\vec \sigma \times \vec k\right)\cdot\hat{y},
\end{equation}
where $\alpha_R$ is the Rashba coefficient, $m^*$ is the effective electron mass, $k$ is the wave vector, and $I_{2\times2}$ is a $2\times2$ identity matrix. The dispersion relation from Eq. \eqref{RashbaH} is given by
\begin{equation}
    E=\dfrac{{\hbar^2k^2}}{2m^*}- s\, \alpha_R {k},
\end{equation}
with $s$ being the spin index. 

Solutions for $k$ for a given energy $E$ are given by
\begin{equation*}
\begin{array}{l}
\hbar^2 k_1 = s\,m^*{\alpha_R} + \sqrt {{\left(m^*\right)^2}\alpha_R^2 + 2\hbar ^2 m^* E}, \\
\hbar^2 k_2 = s\,m^*{\alpha_R} - \sqrt {{\left(m^*\right)^2}\alpha_R^2 + 2\hbar ^2 m^* E},
\end{array}
\end{equation*}
noting that $s^2=1$. 

Here, $k_1(s=+1)$ and $k_1(s=-1)$ correspond to $M$ and $N$ respectively. Similarly $k_2(s=-1)$ and $k_2(s=+1)$ correspond to $M$ and $N$ respectively, staisfying the time-reversal symmetry. Thus the degree of SML $p_0$ is given by
\begin{equation}
%\label{deg_Rashba_SML}
\begin{aligned}
p_0(E_F) &= \dfrac{{k_1}(s=+1)-{k_1}(s=-1)}{{k_1}(s=+1)+{k_1}(s=-1)} \\&= \dfrac{{{\alpha_R}}}{{\sqrt {\alpha_R^2 + \dfrac{{2\hbar^2 E_F}}{m^*}} }}.
\end{aligned}
\end{equation}
which in conjunction with $E_F=\dfrac{1}{2}m^* v_F^2$ gives the expression in Eq. \eqref{RashbaSML}.

\bibliography{main}

\end{document}